\documentclass{aa}                                   
\usepackage{graphicx}
\usepackage[varg]{txfonts}
\usepackage{subfig}
\bibpunct{(}{)}{;}{a}{}{,} 

\def\gtrsim{\mathrel{\hbox{\rlap{\hbox{\lower4pt\hbox{$\sim$}}}\hbox{$>$}}}}
\def\lesssim{\mathrel{\hbox{\rlap{\hbox{\lower4pt\hbox{$\sim$}}}\hbox{$<$}}}}
\newcommand{\hi}{H{\sc i}}

\newcommand{\msun}{M$_{\odot}$}

\begin{document}

\title{The Herschel Virgo Cluster Survey - XIII. Dust in early-type galaxies
\thanks{{\it Herschel} is an ESA space observatory with science instruments provided by European-led Principal Investigator consortia and with important participation from NASA.}}

\author{S. di Serego Alighieri\inst{1}, S. Bianchi\inst{1}, C. Pappalardo\inst{1}, S. Zibetti\inst{1}, R. Auld\inst{2}, M. Baes\inst{3}, G. Bendo\inst{4}, 
E. Corbelli\inst{1}, 
J.I. Davies\inst{2}, T. Davis\inst{5}, I. De Looze\inst{3}, J. Fritz\inst{3}, G. Gavazzi\inst{6}, C. Giovanardi\inst{1}, M. Grossi\inst{7}, L.K. Hunt\inst{1}, L. Magrini\inst{1},  D. Pierini\inst{8} and E.M. Xilouris\inst{9}}

\institute{INAF-Osservatorio Astrofisico di Arcetri, L.go E. Fermi 5, 50125 Firenze, Italy\\
\email{sperello@arcetri.astro.it}
\and School of Physics and Astronomy, Cardiff University, The Parade, Cardiff, CF24 3AA, UK
\and Sterrenkunding Observatorium, Universiteit Gent, Krijgslaan 281 S9, B-9000 Gent, Belgium
\and UK ALMA Regional Centre Node, Jodrell Bank Centre for Astrophysics, School of Physics and Astronomy, University of Manchester, Oxford Road, Manchester M13 9PL, UK
\and European Southern Observatory, Karl-Schwarzschild Str. 2, 85748 Garching bei M\"{u}nchen, Germany
\and Universit\'a di Milano -- Bicocca, Piazza delle Scienze 3, 20126 Milano, Italy
\and CAAUL, Observatòrio Astronòmico de Lisboa, Tapada de Ajuda, 1349-018, Lisboa, Portugal
\and Max-Planck-Institut f\"{u}r Extraterrestrische Physik, Gie\ss enbachstra\ss e, Postfach 1312, 85741 Garching, Germany
\and Institute for Astronomy, Astrophysics, Space Applications \& Remote Sensing, National Observatory of Athens, P. Penteli, 15236, Athens, Greece
}

\date{Received 12 October 2012 / Accepted 10 January 2013}

\authorrunning{S. di Serego Alighieri et al.}
\titlerunning{Dust in ETG with HeViCS}
{}

\abstract{}{We study the dust content of a large optical input sample of 910 early-type galaxies (ETG) in the Virgo cluster, also extending to the dwarf ETG, and examine the results in relation to those on the other cold ISM components.}
{We have searched for far-infrared emission in all galaxies in the input sample using the 250 $\mu$m image of the Herschel Virgo Cluster Survey (HeViCS). This image covers a large fraction of the cluster with an area of $\sim$55 square degrees. For the detected ETG we measured fluxes in five bands from 100 to 500 $\mu$m, and estimated the dust mass and temperature with modified black-body fits.}
{Dust is detected above the completeness limit of 25.4 mJy at 250 $\mu$m in 46 ETG, 43 of which are in the optically complete part of the input sample. In addition, dust is present at fainter levels in another 6 ETG. We detect dust in the 4 ETG with synchrotron emission, including M 87. Dust appears to be much more concentrated than stars and more luminous ETG have higher dust temperatures. Considering only the optically complete input sample and correcting for the contamination by background galaxies, dust detection rates down to the 25.4 mJy limit are 17\% for ellipticals, about 40\% for lenticulars (S0 + S0a) and around 3\% for dwarf ETG. Dust mass does not correlate clearly with stellar mass and is often much greater than expected for a passive galaxy in a closed-box model. The dust-to-stars mass ratio anticorrelates with galaxy luminosity, and for some dwarf ETG reaches values as high as for dusty late-type galaxies. In the Virgo cluster slow rotators appear more likely to contain dust than fast ones. Comparing the dust results with those on \hi\, there are only 8 ETG detected both in dust and in \hi\ in the HeViCS area; 39 have dust but only an upper limit on \hi, and 8 have \hi\ but only an upper limit on dust. The locations of these galaxies in the cluster are different, with the dusty ETG concentrated in the densest regions, while the \hi\ rich ETG are at the periphery.}{}


\maketitle

\section{Introduction}

Dust is a very important component of galaxies, since it is closely connected with their evolution \citep{Dra03}. Dust is thought to be the product of the last stages of the evolution of low- and intermediate-mass stars, of supernovae (SN), and possibly also of active galactic nuclei (AGN). After dust is formed, how much of it survives in the InterStellar Medium (ISM) depends on specific local conditions, and dust can also grow directly in the ISM. The presence of dust has a profound effect on galaxy colours by absorbing blue/UV radiation and re-emitting it in the InfraRed (IR), therefore influencing the amount of escaping ionising radiation. Together with cold gas, dust is associated with regions of star formation (SF), so that dust emission is often used as a tracer for SF. Finally, since dust can be destroyed by the energetic ions in hot haloes and can serve as a catalyzer for the formation of molecules, its presence (or absence) can provide important clues to the relationships between the different phases of the ISM.

Early-type galaxies (ETG) are spheroidal and bulge-dominated galaxies, including ellipticals (E), lenticulars (S0 and S0a), and dwarf spheroidals (dE and dS0). They contain most of the visible mass in the Universe, and are a uniform class of galaxies with a relatively simple morphology and smooth kinematics. They were thought to be devoid of ISM, but in the last decades were found instead to contain a multiform ISM. Its richest component in massive ETG is hot gas, visible in X-rays, but cooler phases are also present, down to the coldest ones, i.e. molecular gas and dust  \citep{Rob91}. In particular the systematic study of the dust content in ETG can provide important clues to the survival of the cold ISM phases in the harsh environment produced by the hot gas (and in some cases by the central AGN), to the presence of residual SF, and to the exchanges of ISM between an ETG and the surrounding InterGalactic Medium (IGM) or other galaxies, throughout its evolution.

The presence of dust in ETG can also change their appearance in subtle ways \citep{Wit92}. For example, colour gradients resulting from dust reddening can mimic those due to metallicity gradients. In addition, \citet{Bae01, Bae02} have suggested that dust scattering can affect the observed kinematics of elliptical galaxies, even mimicking the signature of a dark matter halo. Therefore, it is crucial to have an extinction-independent way to evaluate the amount of dust in ETG, also in order to properly understand their other properties.

The presence of dust in ETG was first inferred from the absorption of stellar light \citep{Ber78, Ebn85, Gou94}. Dust lanes are easy to find on the very smooth light profiles of elliptical galaxies. However, dust absorption does not say much about the quantitative dust content of ETG. Far-infrared (FIR) emission from dust in ETG was later observed with the InfraRed Astronomical Satellite (IRAS) in about 45\% of ellipticals and 68\% of S0 galaxies \citep{Kna89}. More recently \citet{Tem04} detected FIR dust emission with the Infrared Space Observatory (ISO) from about 55\% of a sample of 53 bright ETG, with dust masses ranging from $10^5$ to more than $10^7$ solar masses. Using the Spitzer Space Telescope, \citet{Tem07} found evidence for extended diffuse dust in a similar sample of ETG. Others have used ground-based telescopes to measure the dust emission on a few bright ETG \citep{Lee04, Sav09}. However these observations are limited to relatively small samples of very bright ETG and lack the sensitivity, the resolution, and/or the long wavelength coverage necessary to study dust in large samples of ETG including dwarf galaxies.

\begin{figure*}
\centering
\includegraphics[width=18.0cm]{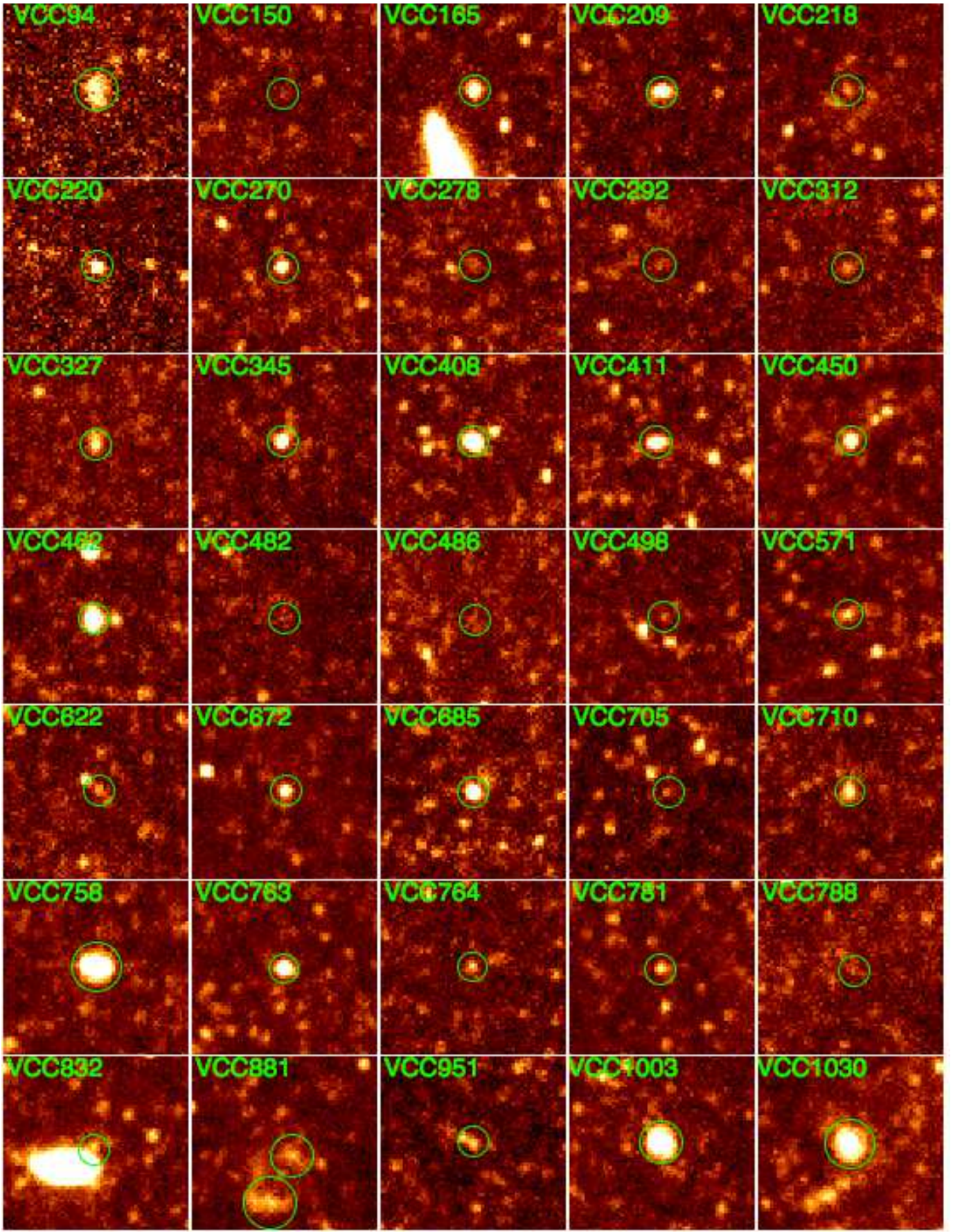} 
\caption{Images at 250 $\mu$m of all the 52 detected ETG. For each image north is up, east to the left, and the size is 6 arcmin. The green circles are the apertures used for flux measurements (continued on the next page).}
\label{fig1a}
\end{figure*}

\begin{figure*}\ContinuedFloat
\centering
\includegraphics[width=18.0cm]{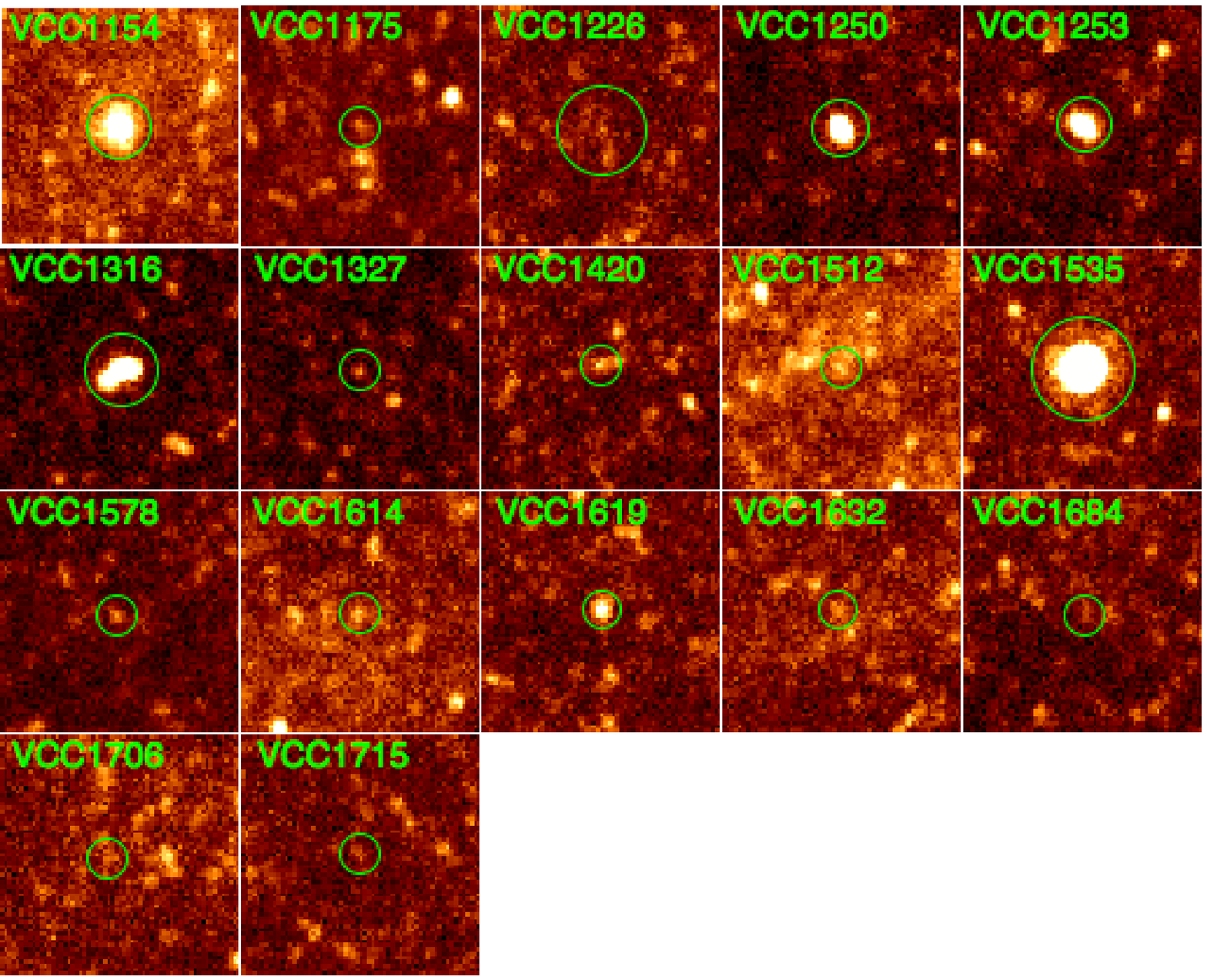} 
\caption{Images at 250 $\mu$m of all the 52 detected ETG (continuation).}
\label{fig1b}
\end{figure*}

The Herschel Virgo Cluster Survey \citep[HeViCS, ][]{Dav10, Dav12}, a confusion-limited imaging survey of a large fraction of the Virgo cluster in five bands --- 100 and 160 $\mu$m with PACS \citep{Pog10} and 250, 350 and 500 $\mu$m with SPIRE \citep{Gri10} --- offers the unique opportunity of carrying out a complete analysis of the dust content of ETG in a nearby cluster.
Auld et al. (2012) have searched the HeViCS images for FIR counterparts of Virgo cluster galaxies of all morphological types with photographic magnitude $m_{pg} \leq 18.0$. The present study instead is limited to ETG, but has no limit on the optical magnitude.
Our work is also complementary to the one by \citet{Smi12} on the ETG of the Herschel Reference Survey \citep[HRS, ][]{Bos10}. This research has searched dust in 62 bright ETG, i.e. a volume-limited sample ($15 Mpc < d < 25 Mpc$) including the Virgo cluster, with $K \leq 8.7$, a bright limit, corresponding to a stellar mass of about $10^{10} M_{\odot}$. Although it is limited to massive galaxies, the HRS sample has the advantage of also including field galaxies. On the other hand, with HeViCS we can also study galaxies which are 1000 times less massive than the HRS limit, but we are limited to the Virgo cluster environment.

In Sect. 2 we discuss the selection of the optical input sample, while in Sect. 3 we describe the methods and the results for the detection and photometry of the dusty ETG. Important quantities are derived from the observed parameters in Sect. 4. Discussions and conclusions are presented in Sects. 5 and 6.
	
\begin{table*}
\caption{The Early-type galaxies detected with HeViCS in the complete FIR detection sample}
\begin{minipage}{\textwidth}
\begin{tabular}{llllclccrrrrr}
\hline \hline
 VCC & Other name & R.A.    & Dec.  & Type & m$_{pg}$ & Dist. & Ap.Rad. & F$_{100}$ & F$_{160}$ & F$_{250}$ & F$_{350}$ & F$_{500}$\\
     &    & h m s   & $^o$ ' "  &  GM\tablefootmark{1} & &  Mpc  & arcsec & mJy & mJy & mJy & mJy & mJy \\
\hline 
94   & NGC 4191  & 12 13 50.40 & +07 12 03.2 & 2  & 13.57 & 32.0 & 42 & $<$300 & 493 & 312 & 156 & 46 \\
165  &           & 12 15 53.28 & +13 12 57.1 & 1  & 14.87 & 17.0 & 30 & 71 & 207 & 167 & 73 & 16 \\
209  & IC 3096   & 12 16 52.36 & +14 30 52.5 & -3 & 15.15 & 17.0 & 30 & 187 & 313 & 221 & 114 & 41 \\
218  & IC 3100   & 12 17 05.55 & +12 17 23.4 & -3 & 14.88 & 17.0 & 30 & $<$32 & 86 & 45 & 13 & $<$15 \\
220  & NGC 4233  & 12 17 07.68 & +07 37 27.8 & 1  & 12.97 & 32.0 & 30 & 192 & $<$178 & 168 & 73 & 35 \\
270  &           & 12 18 06.81 & +05 41 04.3 & 2  & 15.02 & 32.0 & 30 & 252 & 232 & 130 & 46 & 12 \\
278  &           & 12 18 14.42 & +06 36 13.1 & -3 & 15.1 & 23.0  & 30 & 149 & $<$200 & 32 & 11 & 6 \\
292  &           & 12 18 31.72 & +05 50 59.3 & -1 & 17.0 & 32.0  & 30 & $<$32 & 36 & 26 & 23 & 10 \\
312  & NGC 4255  & 12 18 56.15 & +04 47 10.1 & 1  & 13.61 & 32.0 & 30 & 53 & 74 & 59 & 30 & 13 \\
327  &           & 12 19 12.38 & +06 22 53.7 & 1  & 14.8 & 32.0  & 30 & 88 & 130 & 107 & 50 & 18 \\
345  & NGC 4261  & 12 19 23.22 & +05 49 30.8 & 0  & 11.31 & 32.0 & 30 & 294 & 192 & 199 & 179 & 149 \\
408  & NGC 4281  & 12 20 21.52 & +05 23 11.0 & 1  & 12.27 & 32.0 & 30 & 1282 & 1275 & 579 & 217 & 70 \\
411  & NGC 4282  & 12 20 24.31 & +05 34 22.2 & 1  & 14.53 & 23.0 & 30 & 268 & 265 & 201 & 90 & 28 \\
450  &           & 12 21 03.65 & +07 04 39.6 & 1  & 15.08 & 23.0 & 30 & 324 & 309 & 190 & 90 & 29 \\
462  & NGC 4292  & 12 21 16.46 & +04 35 44.5 & 2  & 13.5 & 17.0  & 30 & 851 & 866 & 499 & 199 & 60 \\
482  &           & 12 21 34.08 & +04 46 46.3 & 2  & 14.77 & 17.0 & 30 & $<$32 & 41 & 27 & 24 & 16 \\
498  &           & 12 21 43.31 & +10 14 02.6 & -1 & 18.5 & 17.0  & 30 & 73 & 44 & 26 & 16 & $<$15 \\
571  &           & 12 22 41.16 & +07 57 01.3 & 1  & 14.74 & 23.0 & 30 & 61 & 124 & 96 & 63 & 26 \\
622  &           & 12 23 10.92 & +09 01 43.4 & -1 & 16.7 & 23.0  & 30 & $<$32 & 24 & 35 & 20 & $<$15 \\
672  & NGC 4341  & 12 23 53.56 & +07 06 25.6 & 1  & 14.21 & 23.0 & 30 & 214 & 214 & 133 & 64 & 25 \\
685  & NGC 4350  & 12 23 57.87 & +16 41 36.3 & 1  & 11.99 & 17.0 & 30 & 1536 & $<$1200 & 318 & 114 & 22 \\
705  &           & 12 24 10.96 & +11 56 47.3 & -1 & 17.2  & 17.0 & 30 & $<$32 & 44 & 38 & 22 & 14 \\
710  &           & 12 24 14.52 & +04 13 33.4 & -3 & 14.83 & 17.0 & 30 & 85 & 141 & 119 & 68 & 31 \\
758  & NGC 4370  & 12 24 54.90 & +07 26 41.6 & 1  & 13.69 & 23.0 & 48 & 3071 & 3601 & 1949 & 782 & 252 \\
763  & NGC 4374  & 12 25 03.74 & +12 53 13.1 & 0  & 10.26 & 17.0 & 30 & 932 & 652 & 265 & 146 & 106 \\
764  &           & 12 25 05.61 & +05 19 44.8 & 1  & 14.9 & 17.0  & 30 & 59 & 163 & 73 & 41 & $<$15 \\
781  & IC 3303   & 12 25 15.20 & +12 42 52.6 & -3 & 14.72 & 17.0 & 30 & 165 & 91 & 88 & 51 & 20 \\
788  &           & 12 25 16.83 & +11 36 19.2 & -1 & 15.8 & 17.0  & 30 & $<$60 & $<$58 & 29 & 14 & $<$15 \\
832  &           & 12 25 43.58 & +12 40 27.5 & -1 & 19.0 & 17.0  & 30 & 17 & 49 & 38 & 18 & $<$15 \\
881C & NGC 4406  & 12 26 10.65 & +12 56 23.2 & 0  & 10.06 & 17.0 & 42 & 200 & 282 & 130 & 62 & 24 \\
881SE&           & 12 26 13.64 & +12 54 51.2 &    &      & 17.0  & 51 & $<$54 & 221 & 273 & 168 & 71 \\
951  & IC 3358   & 12 26 54.34 & +11 39 50.3 & -2 & 14.35 & 17.0 & 30 & 179 & 143 & 110 & 54 & 19 \\
1003 & NGC 4429  & 12 27 26.51 & +11 06 27.8 & 2  & 11.15 & 17.0 & 42 & 4436 & 4523 & 2107 & 703 & 198 \\
1030 & NGC 4435  & 12 27 40.49 & +13 04 44.2 & 1  & 11.84 & 17.0 & 48 & 4225 & 3804 & 1897 & 748 & 237 \\
1154 & NGC 4459  & 12 29 00.01 & +13 58 42.1 & 1  & 11.37 & 17.0 & 48 & 4420 & 3738 & 1708 & 617 & 183 \\
1226 & NGC 4472  & 12 29 46.76 & +08 00 01.7 & 0  & 9.31 & 17.0  & 66 & $<$68 & 396 & 86 & 106 & 52 \\
1250 & NGC 4476  & 12 29 59.08 & +12 20 55.2 & 1  & 12.91 & 17.0 & 42 & 1507 & 1435 & 751 & 287 & 103 \\
1253 & NGC 4477  & 12 30 02.20 & +13 38 11.8 & 2  & 11.31 & 17.0 & 42 & 1215 & 1057 & 472 & 178 & 40 \\
1316 & NGC 4486  & 12 30 49.42 & +12 23 28.0 & 0  & 9.58 & 17.0  & 54 & 580 & 825 & 806 & 1033 & 1281 \\
1420 & IC 3465   & 12 32 12.24 & +12 03 41.6 & -1 & 16.41 & 17.0 & 30 & 62 & $<$78 & 73 & 62 & 27 \\
1512 &           & 12 33 34.66 & +11 15 43.0 & -3 & 15.73 & 17.0 & 30 & $<$92 & 157 & 46 & 19 & 8 \\
1535 & NGC 4526  & 12 34 03.03 & +07 41 56.9 & 1  & 10.61 & 17.0 & 78 & 15835 & 15637 & 7992 & 3160 & 1038 \\
1578 &           & 12 34 41.68 & +11 08 34.1 & -1 & 19.7 & 17.0  & 30 & 49 & $<$58 & 29 & 30 & 18 \\
1614 & IC 3540   & 12 35 27.23 & +12 45 00.9 & 1  & 14.44 & 17.0 & 30 & 52 & 89 & 30 & $<$19 & $<$15 \\
1619 & NGC 4550  & 12 35 30.58 & +12 13 15.0 & 0  & 12.5 & 17.0  & 30 & 193 & 306 & 133 & 45 & 6 \\
1632 & NGC 4552  & 12 35 39.81 & +12 33 22.8 & 1  & 10.78 & 17.0 & 30 & 121 & $<$98 & 28 & 30 & 20 \\
1684 & IC 3578   & 12 36 39.41 & +11 06 06.7 & -3 & 14.87 & 17.0 & 30 & 21 & 41 & 30 & 12 & $<$15 \\
\hline \hline
\end{tabular}
\tablefoot{GOLDMine type: -3=dS0 -2=dE$/$dS0 -1=dE(d:E) 0=E-E$/$S0 1=S0 2=S0a-S0/Sa}
\end{minipage}
\end{table*}

\begin{table*}
\caption{Additional Early-type galaxies detected with HeViCS}
\begin{minipage}{\textwidth}
\begin{tabular}{llllccccrrrrr}
\hline \hline
 VCC & Other name & R.A.    & Dec.  & Type & m$_{pg}$ & Dist. & Ap.Rad. & F$_{100}$ & F$_{160}$ & F$_{250}$ & F$_{350}$ & F$_{500}$ \\
     &    & h m s   & $^o$ ' "  &  GM\tablefootmark{1} & &  Mp  & arcsec & mJy & mJy & mJy & mJy & mJy \\
\hline 
150  &           & 12 15 28.57 & +12 38 51.5 & -1 & 20.0 & 17.0 & 30 & $<$32 & $<$18 & 13 & 20 & 6 \\
486  & IC 782    & 12 21 36.97 & +05 45 56.7 & 2  & 14.5 & 23.0 & 30 & 34 & 42 & 18 & 13 & $<$15 \\
1175 &           & 12 29 18.20 & +10 08 09.0 & 0 & 16.01 & 23.0 & 30 & 14 & 38 & 19 & 15 & $<$15 \\
1327 & NGC 4486A & 12 30 57.71 & +12 16 13.3 & 0 & 13.26 & 17.0 & 30 & 155 & 82 & 13 & $<$19 & $<$15 \\
1706 &           & 12 37 11.31 & +12 26 46.1 & -1 & 20.0 & 17.0 & 30 & $<$32 & $<$18 & 23 & 11 & $<$15 \\
1715 &           & 12 37 28.52 & +08 47 40.3 & -1 & 16.2 & 17.0 & 30 & $<$32 & 67 & 21 & $<$19 & $<$15 \\
\hline \hline
\end{tabular}
\tablefoot{GOLDMine type: -3=dS0 -2=dE$/$dS0 -1=dE(d:E) 0=E-E$/$S0 1=S0 2=S0a-S0/Sa}
\end{minipage}
\end{table*}

\section{Sample selection}

The sample of ETG to be searched for dust within HeViCS was selected at optical wavelengths using the GOLDMine compilation \citep{Gav03}, which is mostly based on the Virgo Cluster Catalogue \citep[VCC, ][]{Bin85, Bin93}, including all morphological types from -3 to 2 (i.e. galaxies earlier than S0a-S0/Sa) and excluding galaxies with radial velocity larger than 3000 km/s, since these are background galaxies; we have retained galaxies without a measured radial velocity. With these selection criteria, 925 ETG are within the 4 HeViCS fields. These galaxies constitute our input sample. This includes a large fraction (73.7\%) of all the known Virgo ETG (GOLDMine has 1255 Virgo galaxies with types from -3 to 2 and radial velocity not larger than 3000 km/s), thereby confirming that the HeViCS fields cover most of the Virgo cluster.

We start from an accurate central position for every ETG in the input catalogue, which we obtained mostly from NED\footnote{The NASA/IPAC Extragalactic Database (NED) is operated by the Jet Propulsion Laboratory, California Institute of Technology, under contract with the National Aeronautics and Space Administration.}. However for 287 ETG of our input sample, NED has positions accurate to 25 arcsec, both in R.A. and in Dec., which are based only on the original work of \citet{Bin85}. This accuracy is insufficient for a reliable search of a possible counterpart in the HeViCS mosaic image at 250 $\mu$m, which has a pixel size of 6 arcsec, a point-spread function (PSF) with full width at half maximum (FWHM) of 18 arcsec, and a very high density of objects. We have therefore measured new accurate positions for these 287 objects, using the Gunn $r$-band image of the Sloan Digital Sky Survey (SDSS). We succeeded in measuring 272 (all except 15) with $m_{pg} > 19.0$ (see Appendix A). Therefore, we could reliably look for a FIR counterpart for 910 ETG of our input sample of 925.
Out of the 910 ETG of the input sample, 447 ETG are brighter than the VCC completeness limit ($m_{pg} \leq 18.0$) and form the optically complete part of our input sample.

\section{Methods and results}

\subsection{Detection and photometry}

We searched for a FIR conterpart of the 910 ETG of the input sample in the HeViCS 8-scan mosaic image at 250 $\mu$m. 
We did a blind search for sources in the 250 $\mu$m mosaic, using both the DAOPhot and SussExtractor implementations in HIPE \citep{Ott10}, which produce equivalent results. We then searched the resulting catalogue of sources for those corresponding to the optical positions of the 910 ETG with accurate coordinates in our input catalogue within one pixel (6 arcsec) and with a signal-to-noise ratio greater than 5 ($S/N > 5$). The resulting candidate counterparts were then examined in detail to establish the reliability of the galaxy detection in the 250 $\mu$m image and of its identification with the ETG in the input sample. For the last step we overlayed the 250 $\mu$m image with an optical image of the galaxy. This procedure gave a list of 52 reliable FIR counterparts, all at $S/N > 6$. For all of them the separation between the optical centre of the galaxy and that of its FIR counterpart is $\leq$ 5 arcsec. For these ETG we measured the flux using the aperture photometry package in the Image Reduction and Analysis Facility \citep[IRAF,][]{Tod93} with an aperture of 30 arcsec radius, large enough to include most of the PSF also at 500 $\mu$m, by centering the aperture on the detected counterpart, and by subtracting the sky background evaluated in a concentric annulus with an inner radius of 48 arcsec and an outer radius of 78 arcsec. Table 1 lists the 46 ETG with a 250-$\mu$m flux above our completeness limit of 25.4 mJy. This limit is determined via Monte Carlo simulations to be the flux at which the success rate in recovering artificial point-like sources drops below 95\%. Out of these 46 ETG, 43 belong to the optically complete input sample (i.e. they have $m_{pg} \leq 18.0$), and they represent our complete FIR counterparts out of the complete input optical sample of ETG. The other 6 ETG with an evident counterpart in the 250 $\mu$m image have a 250-$\mu$m flux slightly smaller than 25.4 mJy, and are listed separately in Table 2. All of them are detected in at least one other HeViCS band. 
We remark that the choice of measuring fluxes using an aperture of 30 arcsec in all bands in order to have a consistent SED implies that at 250 $\mu$m this photometric aperture is larger than required for pure detections. Therefore, although all of our 46 counterparts that are brighter than the completeness limit are reliably detected at $S/N > 6$, for some of them the photometric flux measurement is at a lower S/N (see the discussion on photometric errors below).

During the visual examination step we noticed that a few of the detected ETG also had 250 $\mu$m emission outside the 30 arcsec aperture. For these ETG we did the photometry using a larger aperture (see Tables 1 and 2) which includes all the visible FIR flux. For these larger apertures the sky annulus was correspondingly larger. The same apertures and sky annuli were also used for the photometry in the other HeViCS bands for all the 52 ETG detected at 250 $\mu$m. For the SPIRE bands (250, 350, and 500 $\mu$m) we did the photometry on the mosaic images, which include the 4 HeViCS fields, since these handle well objects which fall in the region of overlap between two fields. 
Figure~\ref{fig1a} shows all the 52 ETG detected at 250 $\mu$m.

For the PACS bands (at 100 and 160 $\mu$m) we used the images produced for the most recent HeViCS data release \citep{Aul13}, which represents for our purposes a substantial improvement over previous releases. A low-pass filter is used for the removal of background noise and instrumental drift. Bright sources have to be masked out before applying the filter to avoid producing a flux reduction around these. The most recent release masks out all catalogued Virgo galaxies, as well as all the bright sources present in the data. This process prevents the problem of flux reduction for all the sources in our input catalogue. Since mosaic images are not available for PACS, we performed the photometry on the individual images for each of the four HeViCS fields. 
Although the PACS photometry on the 52 ETG was done blindly using the same aperture used at 250 $\mu$m, every ETG was examined visually in each HeViCS band and flagged in case it was not detected in that band, in which case a flux upper limit appears in Tables 1 and 2.

In the case of VCC 881 (M 86, NGC 4406) we used two circular photometric apertures: one with a radius of 42 arcsec, containing the central dust emission, centred slightly to the south-west of the nucleus (VCC 881C), and one with a radius of 51 arcsec, containing the emission from the filament at about 2 arcmin to the south of the nucleus (VCC 881SE), based on the information on the dust content collected during the HRS Science Demonstration phase \citep{Gom10}.

The methods followed for the photometric error analysis are dictated by the scope of our work, which is aperture photometry of generally faint and compact sources, detected by the use of well-known 
source-extraction algorithms and checked by visual inspection. 
We take into account the stochastic pixel-to-pixel fluctuations in randomly selected sky regions free of bright sources and their effect on our relatively small apertures. These take into account the background error, instrumental noise, and confusion noise. 
In SPIRE on reasonably clean sky areas we measure 1-$\sigma$ pixel-to-pixel fluctuations of 4.9, 4.9, and 5.7 mJy/beam at 250, 350, and 500 $\mu$m, respectively (beam areas are 423, 751, and 1587 arcsec$^2$, and pixel sizes are 6, 8, and 12 arcsec). In PACS we measure pixel-to-pixel 1-$\sigma$ fluctuations of 0.6 and 0.5 mJy/pix at 100 and 160 $\mu$m, respectively (pixel sizes are 2 and 3 arcsec).
Therefore, in our standard 30 arcsec aperture we adopt a 1-$\sigma$ error of 12.7, 9.5, 7.6, 16.0, and 8.9 mJy at 250, 350, 500, 100, and 160 $\mu$m, respectively. For larger apertures we assume that the noise increases as the square root of the number of pixels.
Calibration errors (7\% for SPIRE and 15\% for PACS) are added in quadrature to obtain the total photometric error.

Tables 1 and 2 list the fluxes measured in all five bands and the radius of the circular aperture used consistently in all bands. For the undetected sources an upper limit is given, which is generally equal to twice the photometric error. In noisy regions we give correspondingly higher upper limits.

\subsection{Contamination by background sources}

Since the SPIRE 250 $\mu$m image that we used to search for the FIR counterparts has a high density of background sources, we have estimated the fraction of our counterparts that are probably background galaxies.
We do so in flux bins which are spaced logarithmically by 0.2 (a factor of 1.585). The number of probable background galaxies among our counterparts in the i-th flux bin is:
\begin{equation}
N_{BGi} = n_{BGi} \times N_{opt} \times \pi  r_{max}^{2},
\end{equation}  
where $n_{BGi}$ is the density of background sources that we have estimated from SussExtractor counts in the HeViCS 250 $\mu$m mosaic image, $N_{opt}$ is the total number of optically selected galaxies in our input sample, i.e. 910, and $r_{max}$ is the maximum angular separation between the optical centre of each dust-detected galaxy and the centre of the FIR counterpart, i.e. 5 arcsec (see Sect. 3.1). Table 3 lists the relevant parameters for the background galaxy contamination including the number of detected ETG for each flux bin ($N_{det}$). Since we estimated the density of background sources $n_{BGi}$ from the HeViCS mosaic, its value at higher flux bins contains an increasing fraction of Virgo galaxies. However, this problem does not substantially affect our estimate of the contamination due to background sources since the contamination is relevant only for the lowest flux bins that contain a negligible fraction of Virgo galaxies.

\begin{table}
\caption{Background galaxy contamination at 250 $\mu$m}
\begin{center}
\begin{tabular}{cclr}
\hline \hline
Flux bin & $n_{BGi}$ & $N_{BGi}$ & $N_{det}$ \\
mJy      & $arcsec^{-2}$ & & \\
\hline
25.4 - 40.3 & 6.16 $\times 10^{-5}$ & 4.4 & 11 \\
40.3 - 63.8 & 3.44 $\times 10^{-5}$ & 2.5 &  4 \\
63.8 - 101.1 & 1.10 $\times 10^{-5}$ & 0.78 & 5 \\
101.1 - 160.3 & 2.43 $\times 10^{-6}$ & 0.17 & 6 \\
160.3 - 254.0 & 6.66 $\times 10^{-7}$ & 0.050 & 6 \\
254.0 - 402.6 & 2.22 $\times 10^{-7}$ & 0.016 & 3 \\
402.6 - 638.0 & 1.43 $\times 10^{-7}$ & 0.010 & 4 \\
638.0 - $\infty$ & 2.43 $\times 10^{-7}$ & 0.018 & 7 \\
\hline
25.4 - $\infty$ & 1.11 $\times 10^{-4}$ & 7.90 & 46 \\
\hline \hline
\end{tabular}
\end{center}
\end{table}

The result is that about 8 background galaxies are probably contaminating the 46 detections listed in Table 1; 7 of these are in the two lowest flux bins, which contain 11 dwarf galaxies and 4 lenticulars detected at 250 $\mu$m. Therefore in our Virgo ETG sample between 4 and 8 dwarf galaxies, out of a total of 16, and between 0 and 4 lenticulars, out of a total of 24, are likely spurious FIR detections. Assuming that the contamination rate is roughly independent of m$_{pg}$, the expected number of contaminants in the optically complete sample is about half of the total (447/910). This statistical contamination, however, does not affect the main results of our work.

In the context of possible spurious identifications with background sources, VCC 815 deserves a special mention, since we do detect a FIR source near the position of VCC 815; however, we think that this source is not associated with the VCC galaxy, but with a redder background galaxy. \citet{Bin85} had already noticed this background galaxy to the west of VCC 815, and attributed a redshift of $v_{Hel}$ = 16442 $km/s$ to it, refering to a private communication by Huchra. Indeed, the position of the background galaxy, clearly visible on the SDSS images, coincides exactly with that of the HeViCS 250 $\mu$m detection, while VCC 815 is at about 5 arcsec to the east.

\subsection{Comparison with other work}

\citet{Aul13} have used HeViCS data to study the dust content of the complete part of the VCC, i.e. of those galaxies with $m_{pg} \leq 18.0$, including all morphological types. Of the common complete sample we have detected all of the 42 sources detected by Auld et al. (two are in VCC 881, i.e. M 86). 
In addition, we have also detected 6 ETG of the \citet{Aul13} sample, which they did not detect: VCC 218, 482, 705, 1226, 1420, and 1684. 
Our photometry agrees with that of Auld et al., particularly if one considers that different photometric methods were followed. For the 42 sources detected in both papers the fluxes are consistent within the given errors, while for the sources detected only by us, the measured fluxes are compatible with the upper limits given by Auld et al., except for VCC 705, for which Auld et al. used the coordinates from \citet{Bin85}, which we have found to be inaccurate (see Appendix A). Finally we have detected an additional 5 ETG fainter than $m_{pg} = 18.0$, which were not in the Auld et al. input sample.

\citet{Smi12} have looked at the 62 ETG of the Herschel Reference Survey \citep{Bos10}. The HRS collaboration has obtained Herschel observations of a volume-limited sample (distance between 15 and 25 Mpc) of 323 bright local galaxies ($K \leq 8.7$), including the Virgo cluster. For the galaxies of this cluster the HRS project shares the SPIRE data with HeViCS. Our input sample has 25 ETG in common with the input sample of Smith et al. We detect dust in all of the 9 ETG (2 E and 7 S0) for which Smith et al. detect dust; in addition we detect dust in VCC 1226 (M 49, NGC 4472, HRS 178), VCC 1316 (M 87, NGC 4486, HRS 183), and VCC 1632 (NGC 4552, HRS 211), which Smith et al. consider non-detections. Our photometry for these 3 ETG is consistent with the upper limits of \citet{Smi12}. The fluxes measured in the SPIRE bands for the common sources are generally consistent within the given errors, although they were measured with different photometric methods. Analysing the reasons for the small differences between \citet{Smi12} and this work, we remark that their methods were optimised for bright galaxies and are explained in detail by \citet{Cie12}. They adopt a different photometric method for the point-like sources and for the extended ones. For the first they fit a Gaussian function on the timeline data, while for the latter they do aperture photometry on the reconstructed images, using larger apertures than we did. Although Gaussian fitting of timeline data is likely to give the best S/N ratio for true point sources, we have adopted the same size aperture in all bands and for all objects, because adopting two systematically different methods may compromise photometric uniformity. Moreover, since our sources are galaxies, they may have flux outside the PSF.

\section{Derived quantities}

In this section we estimate additional important physical quantities and describe how they are derived. These depend on the assumed distance: for each galaxy we have used the distance given in GOLDMine (17, 23, or 32 Mpc), which distinguishes various components of the Virgo cluster \citep{Gav99}.

\begin{table*}
\caption{Derived quantities for the Early-type galaxies detected with HeViCS}
\begin{center}
\begin{minipage}{\textwidth}
\begin{tabular}{lcccccccccc}
\hline \hline
VCC & Type & M$_B$ & Dist. & $M_{dust}$ & $T_{dust}$ & $M_{star}$ & Bands & $M_{HI}$ & $M_{H_2}$ & Ref.($H_2$)\tablefootmark{c} \\
    & GM   &  & Mpc & $10^4 M_{\odot}$ & K & $10^8 M_{\odot}$ & for $M_{star}$ & $10^7 M_{\odot}$ & $10^7 M_{\odot}$ & \\
\hline 
94   & 2  & -19.04 & 32.0 & 325$\pm$83 & 19.1$\pm$1.5 & 133$\pm$40 & BVH & 180.2 & $<$9 & Y11 \\
150  & -1 & -11.30 & 17.0 & 13.5$\pm$6.7 & 15.6\tablefootmark{a} & 0.049$\pm$0.013 & gi & $<$3.6 & & \\
165  & 1  & -16.42 & 17.0 & 72$\pm$13 & 16.7$\pm$0.7 & 9.9$\pm$4.4 & BH & $<$7.9 & & \\
209  & -3 & -16.14 & 17.0 & 87$\pm$13 & 17.9$\pm$0.6 & 3.4$\pm$0.9 & giH & 3.9 & & \\
218  & -3 & -16.41 & 17.0 & 20.1$\pm$4.0 & 17.6\tablefootmark{a} & 2.7$\pm$1.8 & BH & $<$3.6 & & \\
220  & 1  & -19.66 & 32.0 & 176$\pm$30 & 19.4$\pm$0.8 & 334$\pm$98 & BVH & $<$27.9 & $<$8 & Y11 \\
270  & 2  & -17.59 & 32.0 & 76$\pm$16 & 22.8$\pm$1.2 & 25$\pm$13 & BH & $<$27.9 & & \\
278  & -3 & -16.81 & 23.0 & 31$\pm$7 & 17.6\tablefootmark{a} & 10$\pm$5 & BH & $<$6.6 & & \\
292  & -1 & -15.60 & 32.0 & 49$\pm$37 & 14.7$\pm$2.3 & 2.3$\pm$0.6 & gi & $<$12.9 & & \\
312  & 1  & -19.00 & 32.0 & 83$\pm$33 & 17.6$\pm$1.5 & 206$\pm$92 & BH & $<$27.9 & $<$6 & Y11 \\
327  & 1  & -17.81 & 32.0 & 142$\pm$35 & 17.7$\pm$0.9 & 7.0$\pm$4.7 & BH & $<$27.9 & & \\
345  & 0  & -21.29 & 32.0 & 64$\pm$5\tablefootmark{b} & 21.3\tablefootmark{a} & 2079$\pm$198 & BVH & $<$27.9 & $<$5 & C07 \\
408  & 1  & -20.34 & 32.0 & 314$\pm$37 & 23.8$\pm$0.9 & 690$\pm$204 & BVH & $<$27.9 & $<$8 & Y11 \\
411  & 1  & -17.35 & 23.0 & 98$\pm$16 & 19.7$\pm$0.8 & 35$\pm$16 & BH & $<$14.4 & & \\
450  & 1  & -16.82 & 23.0 & 82$\pm$14 & 20.9$\pm$0.9 & 4.6$\pm$2.8 & BH & 4.6 & & \\
462  & 2  & -17.74 & 17.0 & 95$\pm$11 & 21.9$\pm$0.8 & 49$\pm$15 & BVH & $<$7.9 & 4.6$\pm$0.4 & Y11 \\
482  & 2  & -16.46 & 17.0 & 32$\pm$22 & 14.6$\pm$2.0 & 13$\pm$4 & BVH & $<$7.9 & & \\
486  & 2  & -17.39 & 23.0 & 9.0$\pm$7.6 & 21.6$\pm$3.8 & 28$\pm$10 & BVH & $<$14.4 & & \\
498  & -1 & -12.76 & 17.0 & 20.8$\pm$6.8 & 15.6\tablefootmark{a} & 0.13$\pm$0.04 & gi & $<$3.6 & & \\
571  & 1  & -17.16 & 23.0 & 110$\pm$27 & 16.1$\pm$0.9 & 6.5$\pm$4.1 & BH & $<$14.4 & & \\
622  & -1 & -15.20 & 23.0 & 30.4$\pm$11.1 & 15.6\tablefootmark{a} & 0.29$\pm$0.08 & gi & $<$6.6 & & \\
672  & 1  & -17.69 & 23.0 & 60$\pm$12 & 20.5$\pm$1.0 & 20$\pm$8 & BVH & $<$14.4 & & \\
685  & 1  & -19.28 & 17.0 & 25.8$\pm$4.5 & 30.9$\pm$2.1 & 237$\pm$77 & BVH & \tablefootmark{e} & $<$1.5 & Y11 \\
705  & -1 & -14.04 & 17.0 & 30$\pm$19 & 15.0$\pm$2.0 & 0.55$\pm$0.15 & gi & $<$3.6 & & \\
710  & -3 & -16.41 & 17.0 & 64$\pm$14 & 16.4$\pm$0.8 & 1.4$\pm$0.7 & BVH & 7.8 & & \\
758  & 1  & -18.22 & 23.0 & 611$\pm$61 & 22.1$\pm$0.7 & 101$\pm$30 & BVH & 8.7 & & \\
763  & 0  & -21.07 & 17.0 & 20.0$\pm$2.7\tablefootmark{b} & 29.0$\pm$1.0 & 1309$\pm$182 & BVH & $<$7.9 & $<$1.7 & Y11 \\
764  & 1  & -16.33 & 17.0 & 30.4$\pm$8.7 & 18.6$\pm$1.1 & 4.9$\pm$1.4 & giH & 6.8 & & \\
781  & -3 & -16.60 & 17.0 & 28.4$\pm$9.4 & 18.4$\pm$1.3 & 5.6$\pm$2.3 & BVH & $<$3.6 & & \\
788  & -1 & -15.49 & 17.0 & 17.5$\pm$7.2 & 15.6\tablefootmark{a} & 1.3$\pm$0.8 & BH & $<$3.6 & & \\
832  & -1 & -12.30 & 17.0 & 19$\pm$12 & 16.4$\pm$2.1 & 0.26$\pm$0.06 & gi & $<$3.6 & & \\
881C & 0  & -21.22 & 17.0 & 29.7$\pm$7.2 & 21.0$\pm$1.2 & 1646$\pm$422 & BVH & 8.9 & & \\
881SE&    &        &      & 249$\pm$56 & 13.9$\pm$0.7 & & & & 2.0$\pm$0.2 & D12 \\
951  & -2 & -16.91 & 17.0 & 27.4$\pm$6.8 & 20.0$\pm$1.1 & 7.2$\pm$2.7 & BVH & $<$3.6 & & \\
1003 & 2  & -20.14 & 17.0 & 228$\pm$23 & 25.3$\pm$0.9 & 682$\pm$178 & BVH & $<$7.9 & 11.2$\pm$0.8 & Y11 \\
1030 & 1  & -19.44 & 17.0 & 266$\pm$28 & 23.9$\pm$0.9 & 250$\pm$83 & BVH & $<$7.9 & 7.4$\pm$0.7 & Y11 \\
1154 & 1  & -19.98 & 17.0 & 189$\pm$20 & 25.8$\pm$1.0 & 413$\pm$104 & BVH & $<$7.9 & 17.4$\pm$0.8 & C07 \\
1175 & 0  & -15.92 & 23.0 & 7.1$\pm$1.9 & 21.3\tablefootmark{a} & 1.8$\pm$1.2 & BH & $<$14.4 & & \\
1226 & 0  & -21.94 & 17.0 & 31$\pm$5 & 21.3\tablefootmark{a} & 2444$\pm$658 & BVH & $<$7.9 & $<$1.8 & Y11 \\
1250 & 1  & -18.37 & 17.0 & 120$\pm$14 & 23.0$\pm$0.9 & 45$\pm$15 & BVH & \tablefootmark{d} & 11$\pm$1 & Y02 \\
1253 & 2  & -19.98 & 17.0 & 57$\pm$8 & 25.3$\pm$1.1 & 353$\pm$104 & BVH & $<$7.9 & 3.5$\pm$0.5 & C07 \\
1316 & 0  & -21.67 & 17.0 & 21.9$\pm$1.8\tablefootmark{b} & 21.3\tablefootmark{a} & 1725$\pm$197 & BVH & \tablefootmark{d} & $<$1.5 & C07 \\
1327 & 0  & -17.99 & 17.0 & 8.4$\pm$1.5 & 21.3\tablefootmark{a} & 113$\pm$22 & BVH & \tablefootmark{d} & & \\
1420 & -1 & -14.87 & 17.0 & 50$\pm$15 & 16.3$\pm$1.2 & 1.2$\pm$0.3 & BVH & \tablefootmark{d} & & \\
1512 & -3 & -15.65 & 17.0 & 19.5$\pm$4.9 & 17.6\tablefootmark{a} & 2.0$\pm$1.2 & BH & $<$3.6 & & \\
1535 & 1  & -20.64 & 17.0 & 1096$\pm$106 & 23.7$\pm$0.8 & 877$\pm$259 & BVH & 1.4 & 39$\pm$1 & C07 \\
1578 & -1 & -11.57 & 17.0 & 27.4$\pm$8.5 & 15.6\tablefootmark{a} & 0.015$\pm$0.004 & gi & $<$3.6 & & \\
1614 & 1  & -16.91 & 17.0 & 8.6$\pm$2.7 & 21.2\tablefootmark{a} & 8.8$\pm$5.0 & BH & $<$7.9 & & \\
1619 & 0  & -18.82 & 17.0 & 25.5$\pm$5.1 & 21.7$\pm$1.0 & 106$\pm$32 & BVH & $<$7.9 & 1.6$\pm$0.5 & C07 \\
1632 & 1  & -20.55 & 17.0 & 8.4$\pm$4.4\tablefootmark{b} & 21.2\tablefootmark{a} & 698$\pm$196 & BVH & $<$7.9 & $<$1.9 & C07\\
1684 & -3 & -16.41 & 17.0 & 10.8$\pm$8.8 & 17.8$\pm$2.9 & 2.2$\pm$0.9 & BVH & $<$3.6 & & \\
1706 & -1 & -11.35 & 17.0 & 13.8$\pm$7.2 & 15.6\tablefootmark{a} & 0.041$\pm$0.011 & gi & $<$3.6 & & \\
1715 & -1 & -15.05 & 17.0 & 21.2$\pm$8.1 & 15.6\tablefootmark{a} & 0.71$\pm$0.46 & BH & $<$3.6 & & \\
\hline \hline
\end{tabular}
\tablefoot{
\tablefoottext{a}{The SED fit is done at fixed temperature.}
\tablefoottext{b}{The SED fit includes a synchrotron component.}
\tablefoottext{c}{Y02 = \citet{You02}; C07 = \citet{Com07}; Y11 = \citet{You11}; D12 = \citet{Das12}.}
\tablefoottext{d}{This source could not be well observed by ALFALFA because of contamination from the M87 radio continuum.}
\tablefoottext{e}{This source is outside the 4-16 deg. declination strip observed by ALFALFA.}
}
\end{minipage}
\end{center}
\end{table*}

\subsection{Dust mass and temperature}

Dust masses and temperatures were estimated by fitting a modifed black-body to the measured HeViCS fluxes for each galaxy. In the case of M86 (VCC 881), separate estimates were performed for the two regions which we measured separately (see Sect. 3).
We used the same procedure as described in \citet{Mag11} \citep[see also][]{Smi10,Dav12}. We assumed
a spectral index $\beta = 2$ and
an emissivity $\kappa [cm^2/g] = 0.192*(350\mu m/\lambda)^2$,
as derived from the Galactic dust emission \citep{Dra03}; we took into account the
filter response function to derive the colour correction. Given
the compactness of our sources, for SPIRE we assumed the filter
response function for point sources; more details and a
table of colour corrections are given in \citet{Dav12}.

We also applied aperture correction factors according to Griffin \citep[private communication, see also][]{Aul13}. For our smallest aperture (30 arcsec radius) these are 1.02, 1.05, 1.13, 1.15, and 1.27 at 100, 160, 250, 350, and 500 $\mu$m, respectively.

\begin{figure*}
\begin{center}
\includegraphics[width=18.5cm]{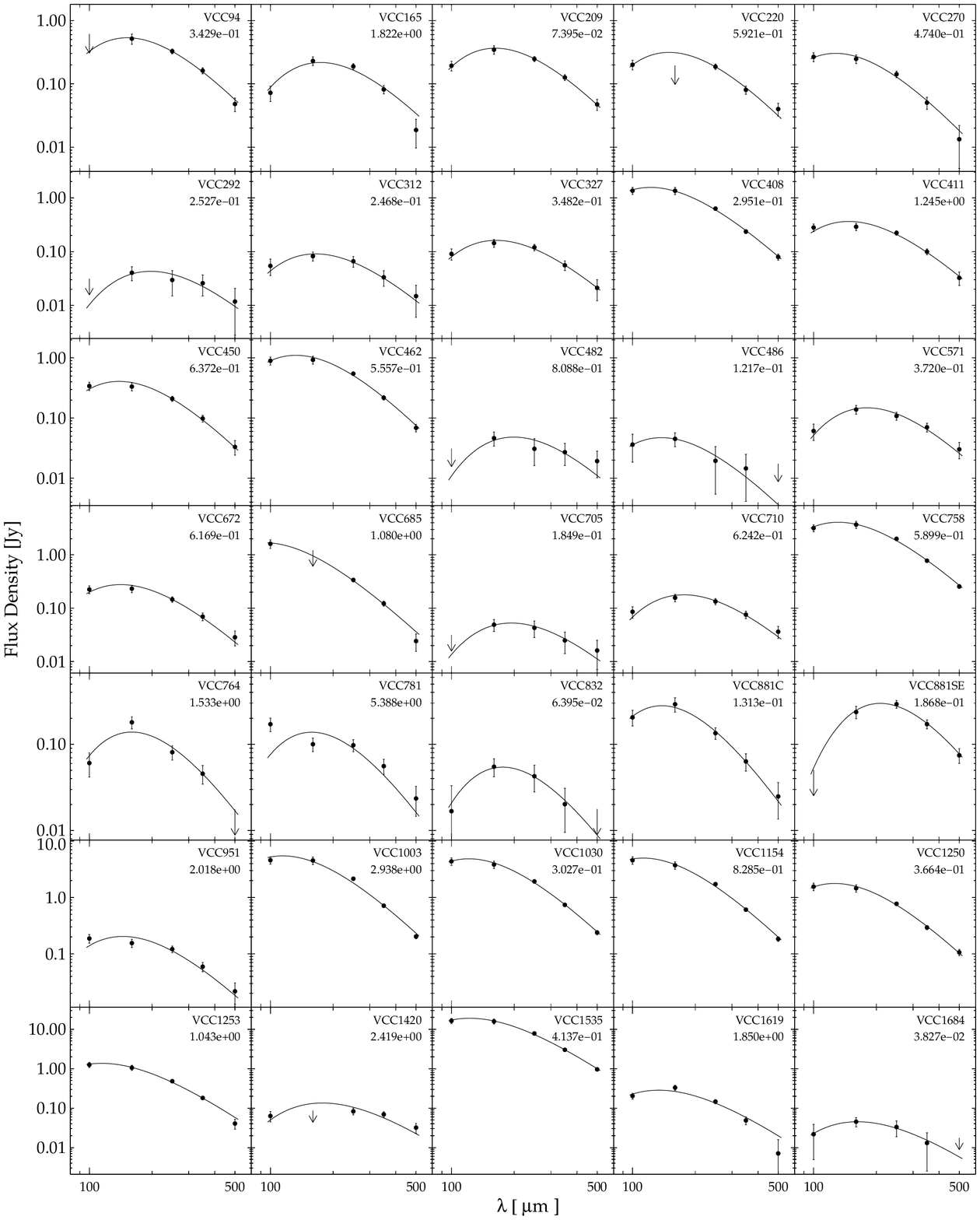} 
\caption{Modified black-body fits to the HeViCS fluxes for all the detected ETG for which fits could be done with temperature as a free parameter and without synchrotron emission. The $\chi ^2$ of each fit is listed under the object name. Upper limits to the fluxes are shown by downward arrows.}
\label{fig2}
\end{center}
\end{figure*}

\begin{figure*}
\begin{center}
\includegraphics[width=18.5cm]{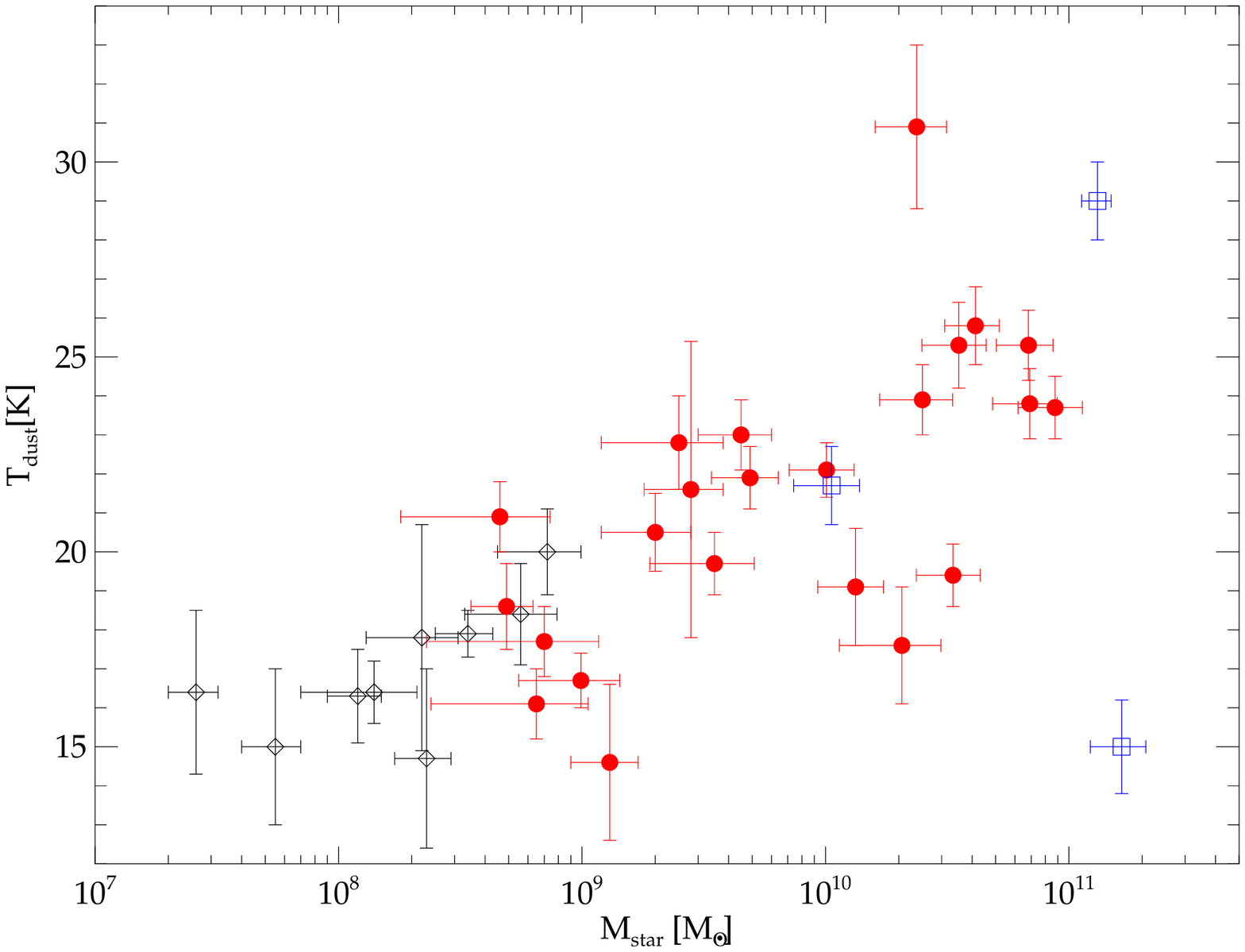} 
\caption{Dust temperature as a function of the stellar mass, divided by morphological type. Open black diamonds are dwarf ETG, open blue squares are ellipticals and filled red circles are lenticulars. Only the 35 ETG with good modified black-body fits are displayed.}
\label{fig3}
\end{center}
\end{figure*}

\begin{figure*}
\begin{center}
\includegraphics[width=18.5cm]{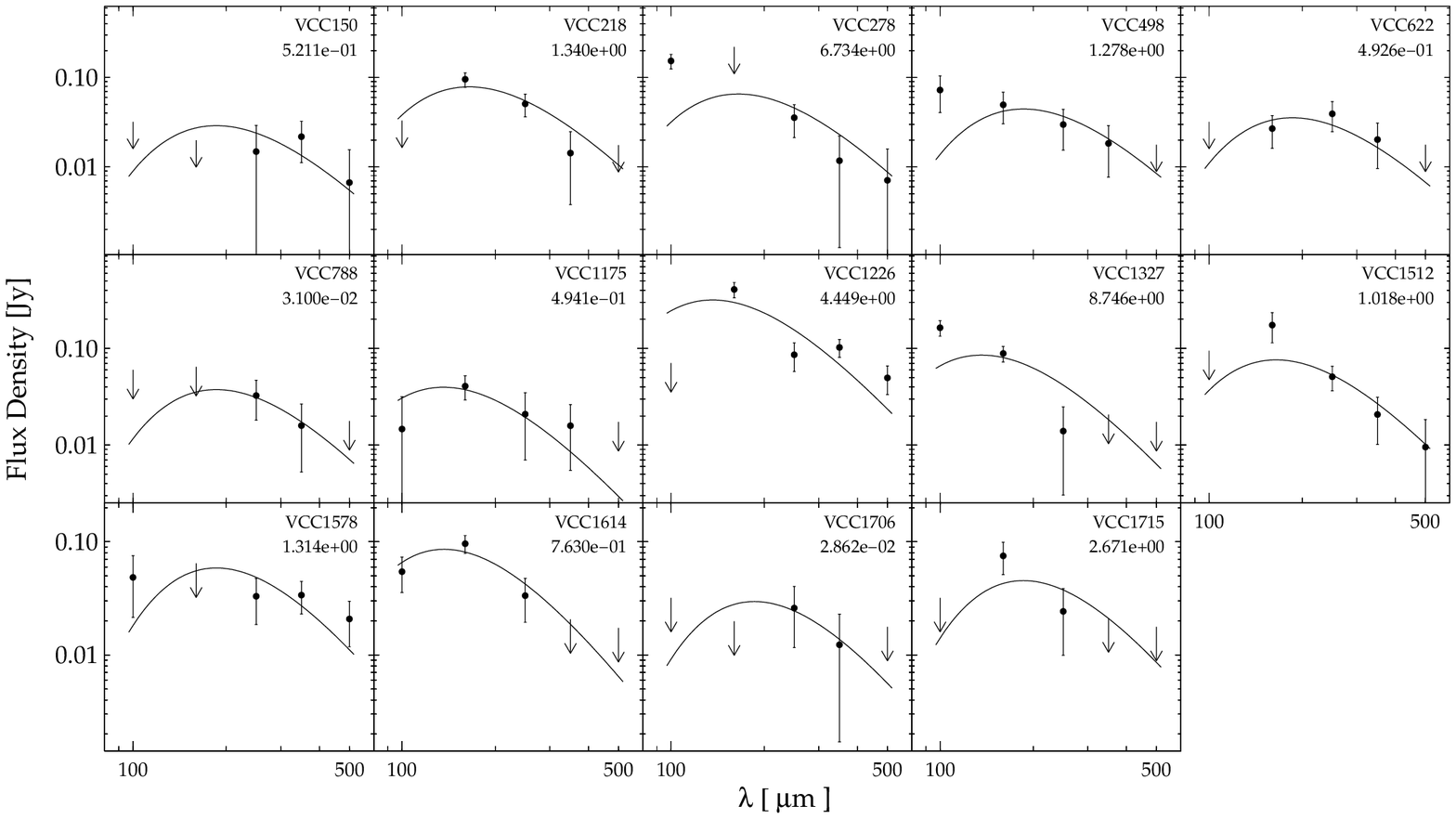} 
\caption{Modified black-body fits to the HeViCS fluxes for the detected ETG fitted with a fixed temperature (see Table 4 for temperature value). The reduced $\chi ^2$ of each fit is listed under the object name. Upper limits to the fluxes are shown by downward arrows.}
\label{fig4}
\end{center}
\end{figure*}

Figure~\ref{fig2} shows the modified back-body fits and Table 4 lists the estimated dust masses and temperatures. For the 14 galaxies for which flux measurements could be made in only 3 bands or fewer, or for which the temperature is not accurately estimated, we have fixed the dust temperature at the average value obtained for the galaxies with the same GOLDMine morphological type (see Fig.~\ref{fig3}). These average temperature values are: $T = 17.6\pm 0.7, 15.6\pm 0.8, 21.3\pm 0.4, 21.2\pm 3.8, 21.4\pm 3.5$ K for GOLDMine types -3, -1, 0, 1, 2, respectively. In these cases the error given for the dust mass in Table 4 takes also into account the variance on the average temperature in the given morphological class. Figure~\ref{fig4} shows the modified black-body fits obtained at the fixed temperature, and results are listed in Table 4.

By using these average dust temperatures, which we have obtained for each morphological class, it is possible to put upper limits to the dust mass of all the 868 Virgo ETG that fall in the HeViCS fields but have not been detected at 250 $\mu$m, by taking into account the upper limit of $F_{250} \leq 25.4$ mJy
\begin{equation}
M_{dust} \leq \alpha \times 25.4 \times 10^4 M_{\odot} \times (Distance/17 Mpc)^2,
\end{equation}
where $\alpha$ = 0.42, 0.70, 0.21, 0.38, 0.34, for GOLDMine types -3, -1, 0, 1 and 2 respectively, including the effect of the variance in the average dust temperature. Dust mass upper limits for the ETG not detected in HeViCS at 17 Mpc, the distance of the main Virgo cloud, range between 5 and $18 \times 10^4 M_{\odot}$, depending on the morphological type, and are correspondingly larger at greater distances.

\begin{figure*}
\begin{center}
\includegraphics[width=18.5cm]{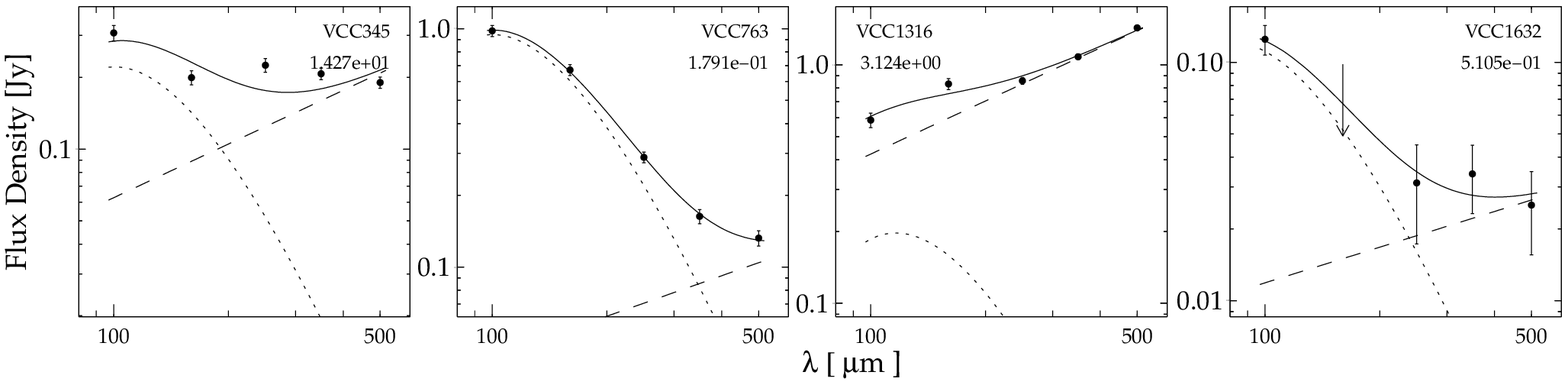} 
\caption{Modified black-body fits to the HeViCS fluxes for the detected ETG with synchrotron emission. The synchrotron component is shown as a dashed line, the dust one as a dotted line. The reduced $\chi ^2$ of each fit is listed under the object name. Upper limits to the fluxes are shown by downward arrows.}
\label{fig5}
\end{center}
\end{figure*}

Four of the ETG detected in HeViCS are known radio sources (VCC 345, 763, 1316, and 1632). The influence of the synchrotron radiation in the HeViCS bands is evident for these objects, particularly at longer wavelengths, and was already seen in \textit{Herschel} data for VCC 763 \citep[M 84, ][]{Bos10b} and VCC 1316 \citep[M 87, ][]{Bae10}. For these objects the fit, in addition to a modified black-body, also includes a synchrotron component, whose slope is fixed at the radio slope (see Fig.~\ref{fig5}). For VCC 345, 1316, and 1632, in the fit we fixed the dust temperature at the average temperature obtained for ETG of the same morphological class (see above), since it could not be reliably determined by the fit itself. All 4 sources show the presence of dust emission, in addition to the synchrotron component; this is also true for M 87, for which our aperture includes both the nucleus and the jet. In M 87 our detection is consistent with the upper limit of \citet{Bae10}, who did not detect dust in the lower quality HeViCS Science Demonstration 2-scan data. We have checked that for each of the four galaxies the amplitude of the fitted synchrotron component agrees well with the radio data, within the uncertainties due to the different apertures used in the radio and in the FIR.

The presence of dust in these radio-loud objects is not completely new, since dust filaments were seen in absorption in the central arcminute of M 87 by \citet{Spa93}, who suggest a dust-to-gas ratio close to the Galactic value.
The presence of dust in M87 was also suggested by \citet{Xil04} by combining their ISO data at 4.5, 6.7, and 15 $\mu$m with IRAS data at 60 and 100 $\mu$m, using an aperture very similar to ours. Their estimate gives a smaller quantity of dust at a higher temperature.
The possible presence of dust near the nucleus of M87 was inferred from spectroscopic observations with Spitzer IRS, which showed an excess emission over the synchrotron component between 30 and 34 $\mu$m, the longest observed wavelengths \citep{Per07}.
An alternative analysis of IRS spectra by \citet{Bus09} in the region up to 24 $\mu$m found no sign of dust emission: their MIR spectrum up to 24 $\mu$m can be completely explained by a combination of a stellar contribution and a synchrotron component.
It is difficult to compare in detail these dust detections with ours for several reasons. First Perlman et al. had a much smaller aperture, about 10 arcsec vs. our 54 arcsec. In addition, they model their excess with a dust component at $T \approx 55 K$, which would contribute less than half of the flux that we have observed at 100 $\mu$m, while our dust component has a negligible contribution to the Spitzer IRS spectrum. It is possible, however, that Xilouris et al. and Perlman et al. have seen warmer dust closer to the nucleus, while we are seeing colder dust over a larger volume. A new analysis of the MIR-FIR-radio SED of M87, including new and deeper Herschel imaging data, is forthcoming (Baes et al., in prep.).

\subsection{Stellar mass}

Since stars are among the major dust producers, it is important to analyse the dust content in comparison with the stellar content, and we have estimated stellar masses for all the dust-detected ETG. 

Stellar masses are estimated using broadband photometry in the optical range
and, where available, in the near infrared (NIR), based on the synthetic libraries and the
approach of \citet[][hereafter ZCR09]{Zib09}.
The mass-to-light ratio M/L as a function of different colours or pairs of colours is obtained
from the 50000 models of the ZCR09 library, which are built based on
the 2007 version of the \citet{Bru03} models, assuming a \citet{Cha03} IMF.

Following ZCR09, the most reliable stellar mass estimates are obtained
using a pair of one optical colour (e.g. $g-i$, $B-I$, $B-V$) and one
optical-NIR colour (e.g. $i-H$, $V-H$) to compute M/L in a NIR band
(e.g. $H$). We can apply this method to 30 galaxies in our sample. For
28 galaxies deep $B$-, $V$- and $H$-band photometry ($BVH$) is provided by the
GOLDMine compilation, with total magnitudes determined from growth curves. Another
2 galaxies have GOLDMine $H$-band, but we have to rely on the SDSS photometry for
the optical bands, namely $g$ and $i$ ($giH$ in Table 4). In this case total GOLDMine
magnitudes are combined with total ``model'' magnitudes of the SDSS,
thus with larger systematic uncertainties due to a different
extrapolation method.
For the remaining galaxies only one colour is used to estimate M/L.
Fourteen galaxies have deep $B$- and $H$-band photometry ($BH$) listed in GOLDMine: in this case the $B-H$ colour is used
to estimate M/L in the $H$-band\footnote{The small bias introduced by the lack
of a second optical band is corrected using the empirical correlation
between $B-H$ and the difference in $\log M/L$ based on $BVH$ and on $BH$
only for the 28 galaxies where the three bands are available.}. Another 8 galaxies do
not have any photometry available from GOLDMine because of their low
luminosity, so model magnitudes in
$g$ and $i$ are extracted from the SDSS catalogs and stellar masses
are computed based on the $i$-band luminosity ($gi$ in Table 4). Although this
combination of bands was shown to provide accurate results \citep[ZCR09,][]{Tay11} 
similar to the two-colour method, stellar masses
for this subsample should be regarded as uncertain owing to the
large photometric errors.
All magnitudes are corrected for foreground Galactic
extinction using the optical depth derived from \citet{Sch98} and 
assuming the standard extinction laws of \citet{Car89} and \citet{ODo94}.

The stellar masses for the dust-detected ETG are listed in Col. 7 of
Table 4, with errors accounting \emph{only} for the intrinsic M/L
scatter at fixed colour(s), as derived from the model library. While
this is the largest contribution to the error budget for the brightest
galaxies, fainter galaxies may indeed have much larger uncertainties
because of photometric errors. The combination of bands used for the
mass estimate is listed in Col. 8.

\subsection{Other ISM components}

\begin{figure*}
\begin{center}
\includegraphics[width=19.0cm]{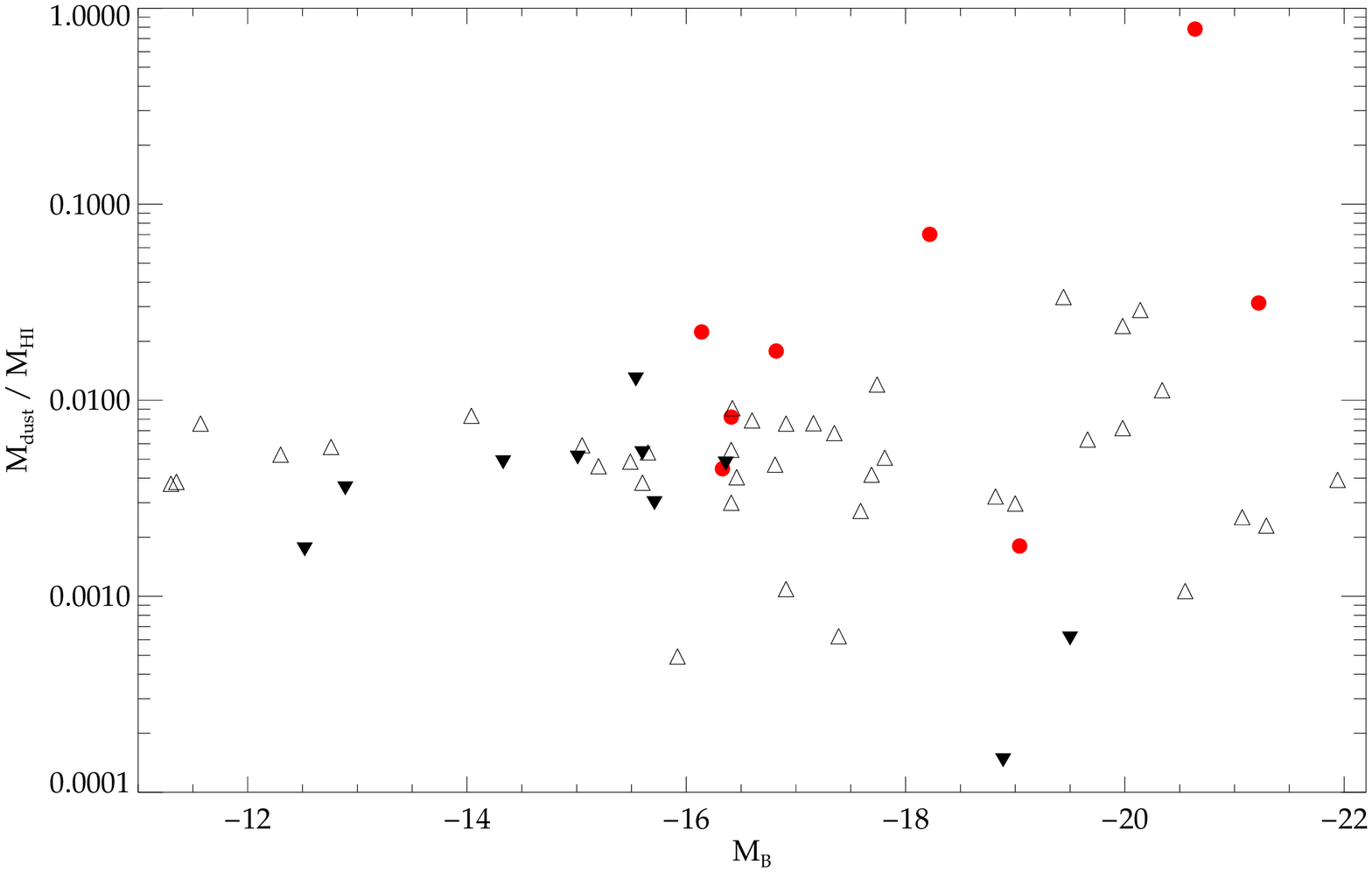} 
\caption{The dust-to-gas mass ratio as a function of the absolute $B$-band magnitude for ETG in the Virgo cluster. Upward open black triangles are \hi\ mass upper limits for dust-detected ETG, while downward filled black triangles are dust mass upper limits for \hi -detected ETG.}
\label{GasToDust}
\end{center}
\end{figure*}

Table 4 also contains complete information about the neutral atomic hydrogen (\hi) content of the ETG detected in HeViCS. This information is obtained from the \hi\ survey of ETG in the Virgo cluster done by \citet{diS07}, using the Arecibo Legacy Fast ALFA \citep[ALFALFA,][]{Gio05, Gio07}. Appendix B of the present paper contains an update of that survey to extend it to the 4-8 deg. declination strip, which has recently become available in ALFALFA \citep{Hay11}. The result is that the HeViCS fields are now almost completely covered by the ALFALFA survey. Table 4 also contains the \hi\ mass for VCC 450 and VCC 758, which have been detected in \hi\ in the deeper Arecibo survey AGES \citep{Tay12}. The upper limits are described in \citet{diS07} for an assumed distance of 16.7 Mpc to the whole of the Virgo cluster, and are given in Table 4 for the distances assumed in this paper.
Figure~\ref{GasToDust} shows the dust-to-atomic-gas  mass ratio for Virgo ETG, showing large variations, even for objects with the same luminosity.

We also give in Table 4 the information about the molecular gas (H$_2$) content which is available in the literature. This information, contrary to that for dust and \hi , is far from complete for the ETG of our input catalogue. The only ETG in our sample that has been detected both in atomic and molecular gas, i.e. the S0 galaxy VCC 1535 (NGC 4526), has a high molecular-to-atomic-gas mass ratio ($M_{H_2}/M_{HI}$ = 28), consistent with the trend to higher values in this ratio for earlier spirals \citep{You91}, but considerably higher than the values reported by \citet{Wel10} for about a dozen elliptical and lenticular galaxies.
In summary only one ETG in our sample (VCC 1535) has been detected in \hi , H$_2$, and dust, while 8 ETG have been detected in both \hi\ and dust, and 9 ETG in H$_2$ and dust. For these 9 galaxies the dust-to-molecular-gas mass ratios are all between $1\times 10^{-2}$ and $4\times 10^{-2}$, and the lower limits to this ratio are consistent with this range.

\section{Discussion}

\subsection{Dust detection rates}

\begin{figure}
\begin{center}
\includegraphics[width=9.0cm]{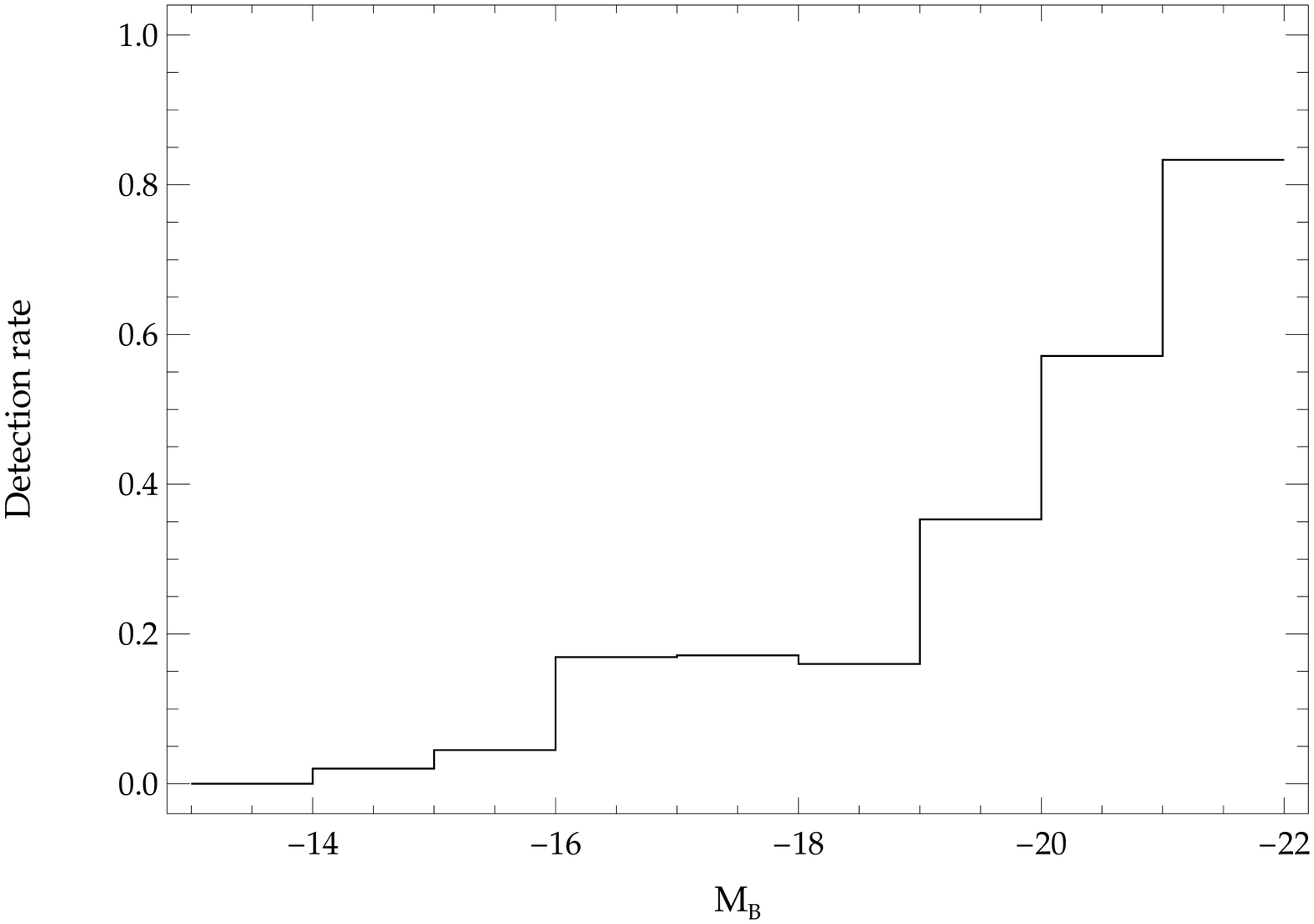} 
\caption{The dependence of the dust detection rate on the optical luminosity for our complete samples.}
\label{DetRate_MB}
\end{center}
\end{figure}

Since detection rates are meaningful only on complete samples,
we examine detection rates only for the 43 galaxies detected above the FIR completeness limit ($F_{250} \geq 25.4$ mJy), and in the optically complete input sample ($m_{pg} \leq 18.0$). For the dwarf ETG (GOLDMine types = -3, -2, -1) we detect 13 galaxies out of 354 (3.7\%); for the ellipticals (GOLDMine type = 0) we detect 6 out of 35 galaxies (17.1\%); and for the lenticulars (GOLDMine types = 1 and 2) we detect 24 galaxies out of 58 (41.4\%). The detection rates for ellipticals and lenticulars are lower than the corresponding rates obtained for the Virgo part of HRS sample by \citet{Smi12} which are 29\% and 53\%, respectively, and are also lower than those obtained by \citet{Kna89} with IRAS and by \citet{Tem04} with ISO. The reason for our lower detection rates is probably because our input sample includes fainter objects, and the detection rate increases with the optical luminosity, as shown in Fig.~\ref{DetRate_MB}.
On the observational side, the dependence of the detection rate on optical luminosity could be related to our dust detection limits, which favour more luminous galaxies. On the other hand, it might also be related to the difficulties that dwarf galaxies have in retaining their dust, particularly if the dust-to-star mass ratio reaches such high values to see the onset of massive star formation, which would lead to SN feedback and to a fast depletion of the ISM.

There are two remarks about our detection rates, however. The first is due to the fact that our complete input sample, which we selected to avoid galaxies with $V_{hel} > 3000$ km/s, actually includes 102 ETG for which the radial velocity is not known (these are all dwarf ETG). Potentially, some of these can have a radial velocity larger than 3000 km/s. They would then be background galaxies not belonging to the Virgo cluster and should be excluded from our input sample\footnote{However \citet{Bin85} have classified, based on their visual inspection, all of these 102 ETG without radial velocity either as cluster members (95) or as possible members (7).}.
By assuming that the probability for a galaxy to be a background contaminant is independent of whether a measured radial velocity is available or not, the fraction of contaminants among VCC galaxies without a valid radial velocity measurement is the same as among VCC galaxies with available radial velocity, i.e. 7.8\%. Hence, only 8 of the 102 ETG without a measure of $V_{hel}$ in the optically complete sample are expected to be background contaminants, and should therefore be excluded from the complete input sample.
Assuming that these background galaxies are among those not detected in dust (in fact only two ETG of the complete detected sample for the optically complete input sample have no measured radial velocity), then the dust detection rate for dwarf ETG becomes 3.8\% (13 out of 354-8=346). 

The second remark is that in Sect. 3.2 we examined the probability that some of our FIR counterparts are actually associated with background galaxies, whose position coincides with that of a Virgo ETG of our input sample, and concluded that these could be 4-8 dwarf galaxies misidentifications, and 0-4 lenticular ones. In this case then, the dust detection rate for dwarf Virgo ETG would be reduced to between 2.6\% and 3.2\%, and the rate for lenticular galaxies to between 38\% and 41\%. The dust detection rates for the brighter ETG have not changed, since all of them have a measured radial velocity, and are not likely to be confused with background sources.

\subsection{Dust vs. stars}

\begin{figure*}
\begin{center}
\includegraphics[width=19.0cm]{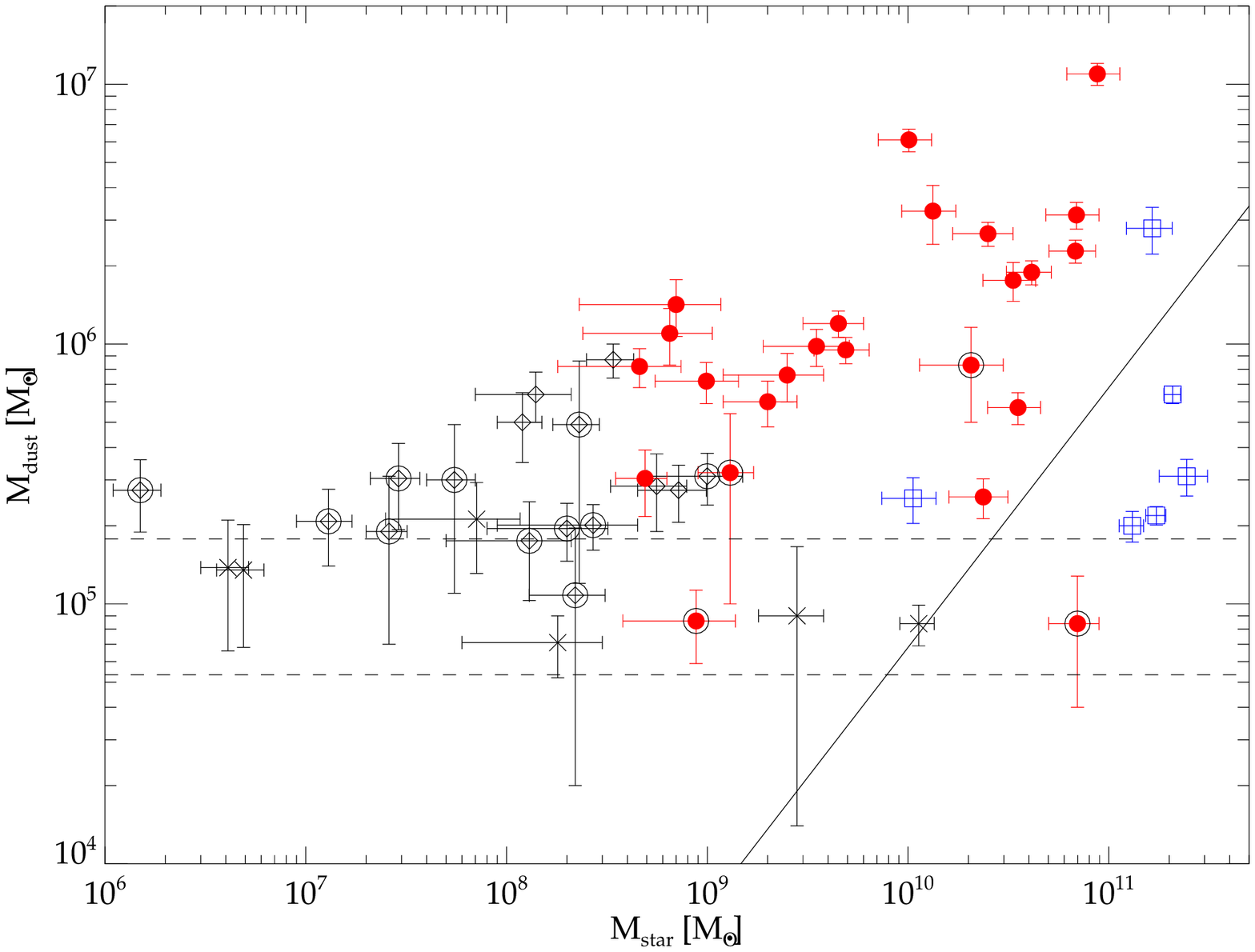} 
\caption{Dust mass as a function of the stellar mass, divided by morphological type. Open black diamonds are dwarf ETG, open blue squares are ellipticals and filled red circles are lenticulars. Black crosses are objects which have a 250-$\mu$m flux below the limit of 25.4 mJy. Objects with a black circle are in the two lowest flux bins of Table 3: half of them are probably contaminating background galaxies. The continuous line shows the dust mass upper limit for the closed-box passively-evolving model (Eq. 3). The dotted lines show the range of the dust mass detection limit for a distance of 17 Mpc (Eq. 2).}
\label{MdustMstar}
\end{center}
\end{figure*}

Dust could originate from stellar/SN mass loss within each galaxy, or it could have an external origin. If the former were true, we should see a dependence of the dust mass on the stellar mass, at least within each morphological type, assuming similar star-formation histories and similar efficiencies for the processes of dust destruction, depletion and loss. Figure~\ref{MdustMstar} shows the comparison between dust mass and stellar mass. There is no clear dependence, particularly if one takes into account that the lower part of the figure is empty because of the dust detection limits, which vary between 5 and 18 $\times 10^4$ \msun \  at 17 Mpc, and are larger at greater distances (see Sect. 4.1). \citet{You11} also found no dependence of the molecular gas content of Virgo ETG with their mass. The dust-detected ETG span about 6 orders of magnitude in stellar mass, but only 2.5 orders of magnitude in dust mass.

We now examine the idea that the dust produced locally in a passive galaxy can account for the observed dust content.
Assuming that an ETG was already in place at z=1 and its stellar
population has passively evolved since z=10 as a closed-box single-stellar-population model \citep[Solar metallicity, Salpeter IMF for 0.1-120 \msun ,][]{Pie04}, about 10\% of its mass was returned by stellar mass loss since z=1. The
dust mass associated with these stellar outflows that survives today should at most give a dust-to-star
mass ratio of
\begin{equation}
{\frac{M_{dust}}{M_{star}}} \leq \frac{1}{9} * {\frac{M_{dust}}{M_{HI}}} * {\frac{t_{surv}} {t_{z=1}}} = 0.11 * \frac{1}{150} * {\frac{7.1\times 10^7} {7.7\times 10^9}} = 6.8\times 10^{-6},
\end{equation}
where for the dust-to-neutral-gas mass ratio we have used the Galactic value, and for the dust grain survival time we have used the constraint $t_{surv} < 46+25$ Myr, the maximum allowed by \citet{Cle10} for a passive galaxy.
The constraint given above on the dust-to-star mass ratio is satisfied by our observations of the Virgo ETG sample only
for a few very massive ETG (see Figs. 8 and 9). On the other hand, it is based on rough
numbers and simple assumptions. Large discrepancies from analogous values
obtained from the previous equation along the mass range should be
understood as differences in the origin of dust (in general: ISM,
outflows from SN Type II/OB associations, and dust cycle in diffuse media,
or dust of external origin) or higher values of dust-to-neutral-gas mass ratio or of the dust grain survival time, for instance.

\begin{figure*}
\begin{center}
\includegraphics[width=19.0cm]{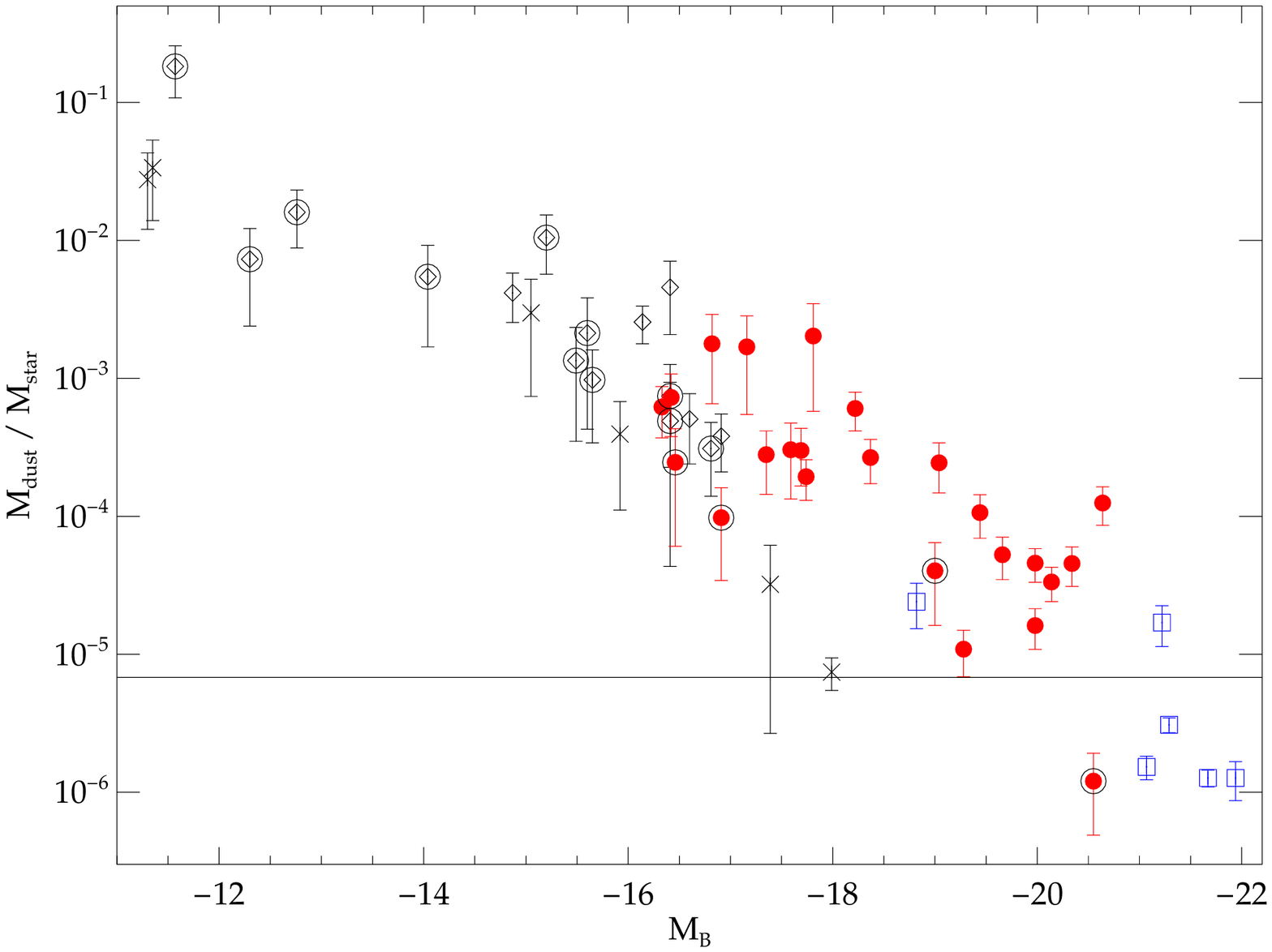} 
\caption{The dust-to-star mass ratio as a function of $B$-band luminosity divided by morphological type. Symbols are as in Fig. 8. The continuous line shows the upper limit for the closed-box passively-evolving model (Eq. 3).}
\label{MdustMstarMB}
\end{center}
\end{figure*}

Looking at the dust-to-stars mass ratio shown in Fig.~\ref{MdustMstarMB}, there is a clear tendency of this quantity to decrease with luminosity for the dust-detected ETG, an effect which cannot be totally due to our incompletenesses. The observational limits on the dust detection can account for the lack of objects in the lower left of the figure, but not in the upper right. While ellipticals and S0 have dust-to-star mass ratios between $10^{-6}$ and $10^{-3}$, this ratio can raise to about half a percent for dwarf ETG (excluding higher values obtained for objects possibly contaminated by background galaxies), as much as for the dusty late-type galaxies \citep{Cor12}. Interestingly, the lenticular galaxies span a range of dust-to-star mass ratio which goes to substantially larger values than the HRS sample \citep{Cor12,Smi12}. This is due to the presence of fainter galaxies of this type in our sample.

\begin{figure}
\begin{center}
\includegraphics[width=9.0cm]{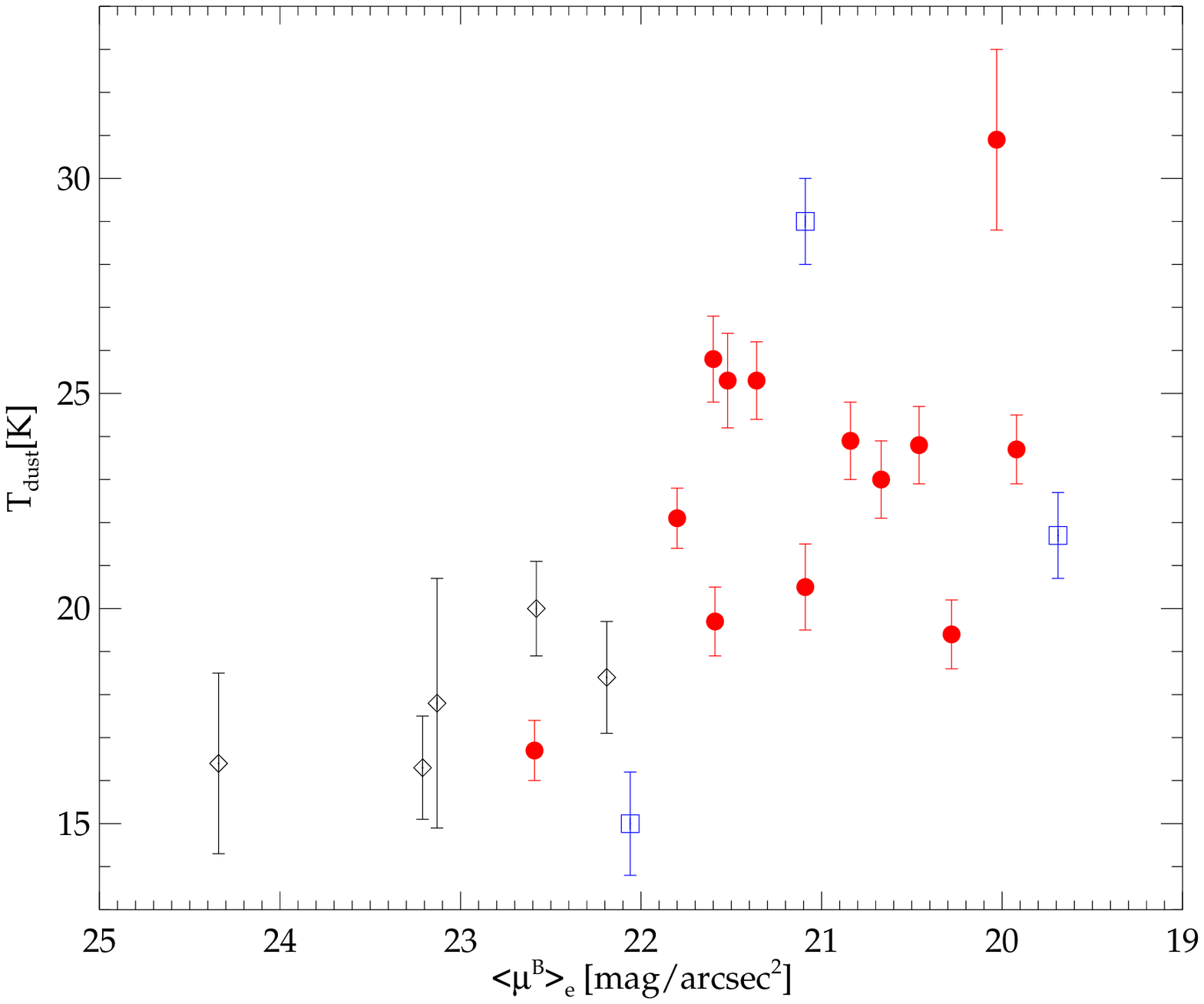} 
\caption{The correlation between the dust temperature and the $B$-band average surface brightness within the effective radius for the object for which this radius is available. Open black diamonds are dwarf ETG, open blue squares are ellipticals and filled red circles are lenticulars.}
\label{TdustMue}
\end{center}
\end{figure}

Figure~\ref{fig3} shows that the dust temperature correlates with the stellar mass of the galaxy. ETG with stellar masses up to $10^9$ \msun \  exhibit cold dust temperatures ($T_{dust} \lesssim 20$K in agreement with values that are typical
of the diffuse interstellar medium in star-forming galaxies. On the other
hand, temperatures tend to be higher in more massive ETG. The latter galaxies
might have
an increased contribution to the emission at the PACS wavelengths from
circumstellar dust or from stochastic heating, or a more intense interstellar
radiation field, associated with post-main-sequence phases of the evolution of
super-solar metallicity low-intermediate-mass stars or, alternatively,
recent star formation activity \citep[e.g.][]{Kav07}. The higher dust temperature in more massive galaxies could also be due to conductive heat transfer from the X-ray gas \citep[][and references therein]{Spa12}.
Figure~\ref{TdustMue} shows that the dust temperature also correlates with the $B$-band average surface brightness within the effective radius, as listed in GOLDMine, indicating clearly a higher dust heating in objects with a stronger radiation field.

We find that dust is much more concentrated than stars. In most of the dust-detected ETG the FIR counterpart is broadly consistent with a point source at 250 $\mu$m. The few clearly extended counterparts are for the 12 galaxies for which we had to use a photometric aperture larger than 30 arcsec, as listed in Table 1 and shown in Fig.~\ref{fig1a}, but also in these cases the dust tends to be much more concentrated than stars.

The ATLAS$^{3D}$ collaboration \citep{Cap11}\footnote{ATLAS$^{3D}$ is a project for obtaining two-dimensional kinematics for a volume-limited sample of 260 bright ETG ($D < 42$ Mpc, $M_K < -21.5$).} convincingly argues that the most important distinction among (bright) ETG is between slow and fast rotators \citep{Ems11}. Therefore, although most of the ETG of our input sample lack the detailed kinematic information necessary for this distinction, we have looked at the relationship between the presence of dust and the stellar kinematics for those objects where this is possible. We note that among the 50 ETG that are in common between our input sample and the ATLAS$^{3D}$ sample (and therefore have accurate kinematics), 13 are classified as slow rotators within one effective radius, and 37 are classified as fast rotators \citep{Ems11}. Since we detect dust above the completeness limit of 25.4 mJy at 250 $\mu$m in 9 slow rotators ($69\pm 23$\%) and in 9 fast rotators ($24\pm 8$\%)\footnote{Two additional slow rotators (IC 782 and NGC 4486A) are detected below the completeness limit (see Table 2), but are not used in these statistics.}, it appears that the slow rotators, although less abundant in a cluster, are much more likely to contain dust. This result surprisingly contrasts the tendency for molecular gas to be found preferentially in fast rotators: \citet{You11} in the ATLAS$^{3D}$ sample find that the CO detection rate is $6\pm 4$\% among slow rotators and $24\pm 3$\% among fast ones. 

This difference suggests that the relationship between molecular gas and dust on the one hand, and kinematics on the other hand might depend strongly on the environment, being different in the Virgo cluster and for field galaxies. Most of the ATLAS$^{3D}$ galaxies, i.e. those for which \citet{You11} obtained their results on the molecular gas, are outside Virgo. In addition, if one takes only the 50 ETG that we have in common with ATLAS$^{3D}$, molecular gas is detected in 23\% of the slow rotators (3 out of 13) and in 14\% of the fast ones (5 out of 37), with slow rotators more efficiently detected, similar to the rates we find for the dust detections, and contrary to the molecular gas detection rates found in the whole ATLAS$^{3D}$ sample, which is mostly made of field objects.
However, the difference could be caused by the presence of kinematically peculiar objects among the dusty slow rotators in the Virgo cluster, such as galaxies with counter-rotating components mimicking slow rotation. Nevetheless this possibility can be excluded, since the brightest ellipticals and lenticulars in the Virgo cluster like M49, M84, M86, M87, M89, NGC 4261 and NGC 4552 are among the dusty slow rotators, and only one dusty slow rotator (NGC 4550) has counter-rotating components. This strengthens our suggestion that the ETG most likely to contain dust and molecular gas are slow rotators in the Virgo cluster and fast rotators in the field. \citet{Dav11} find that the gas has a purely internal origin (and a smaller amount) in fast rotators residing in dense environments, while it has an external origin, possibly leading to larger amounts, in about half of the fast rotators in the field. They also find that the dominant source of gas is external for slow rotators, which are mainly found in dense enviroments.
We tentatively conclude that the cluster environment might favour cold ISM accretion from other galaxies particularly for slow rotators, which appear to be more concentrated in the densest parts of the cluster than fast rotators.

\subsection{Dust vs. gas and spatial distributions}

Although dust detection rates and atomic-gas detection rates \citep[see][and Appendix B]{diS07} are similar for the Virgo ETG, and the dust-to-gas mass ratios which we obtain are similar to those obtained by \citet{Cor12}, the comparison between Table 4 and Table B1 in Appendix B (see also Fig.~\ref{GasToDust}) shows that there is a surprisingly small overlap between dust-detected and atomic-gas-detected ETG: there are only 8 ETG detected both in dust and in atomic gas, while there are 39 ETG detected in dust, but not in atomic gas, and 8 ETG detected in atomic gas, but not in dust. This comparison has a strong observational basis, given the completeness of both the atomic gas and the dust surveys for Virgo ETG.

\begin{figure*}
\begin{center}
\includegraphics[width=19.0cm]{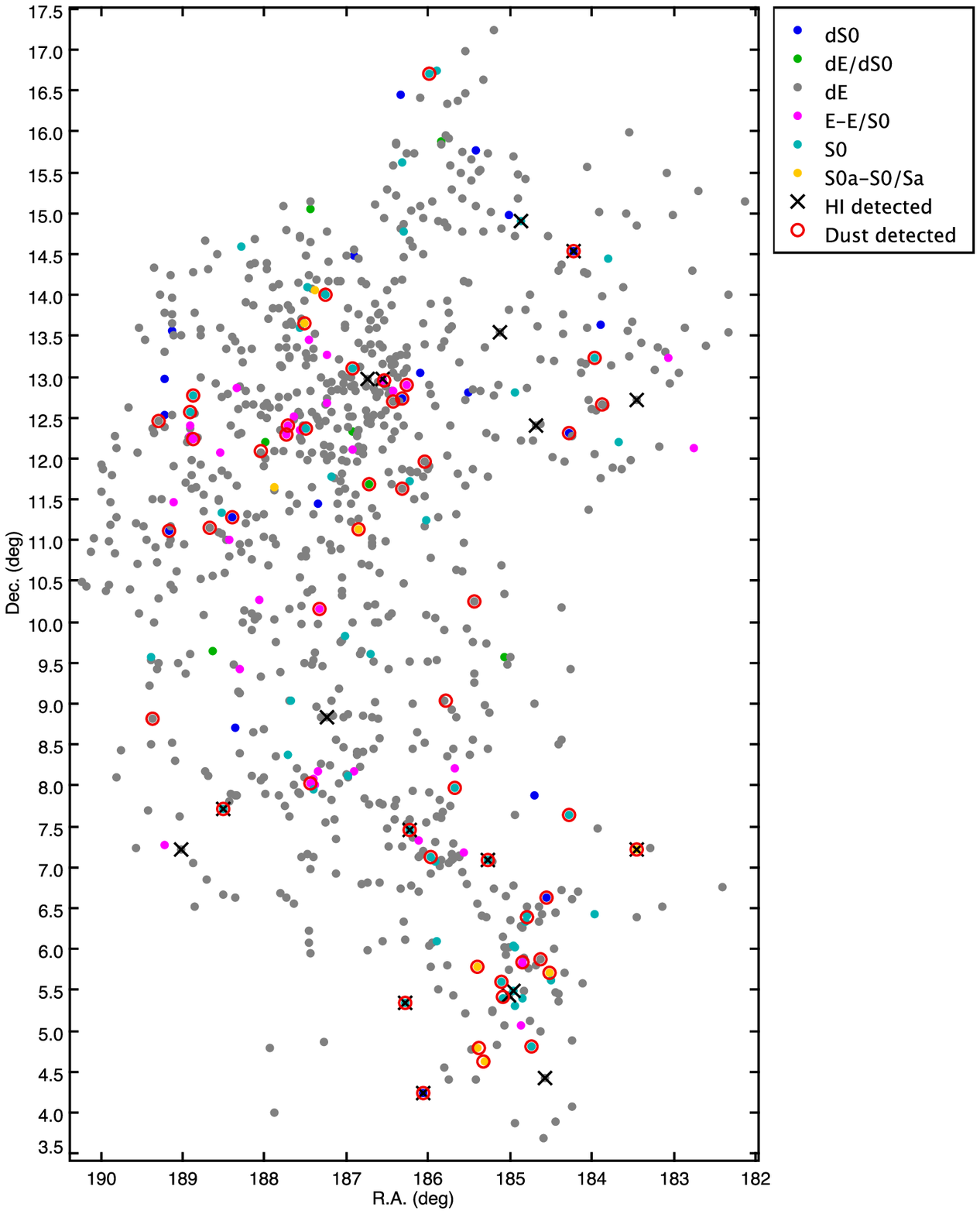} 
\caption{Map of all the ETG examined in HeViCS, distinguished by morphological type. Those marked with a red circle have been detected in dust, while those marked with a black X have been detected in \hi .}
\label{Map}
\end{center}
\end{figure*}

An additional important clue to this incompatibility between dust and atomic gas in Virgo ETG is given by their respective location in the cluster.
Figure~\ref{Map} illustrates the spatial distribution in the Virgo cluster of all the ETG examined in this work and of those for which dust and \hi\ have been detected (see Appendix B). These spatial distributions show several interesting features. First, the dust-detected ETG tend to concentrate at the centre of the main clouds in the Virgo cluster, i.e. the central cloud A around M87, cloud W to the south-west, and cloud M to the west, while, as noticed by \citet{diS07}, the \hi -detected ETG tend to be at the periphery of the cluster. This fact remains true even considering the \hi -blind region within one degree from M87. We suggest that this is because atomic gas is more affected than dust by the hot gas, which tends to concentrate on the densest regions of the cluster. This would have implications on the star-formation rate of cluster ETG, in particular if, as we suspect, the distribution of molecular gas is more similar to that of the dust than to that of atomic gas.
The tendency of dust-detected ETG to concentrate on the cluster clouds is reinforced by the fact that 15 out of the 25 lenticular galaxies present in the south-west part of the cluster (RA $<$ 186.5 deg. and Dec. $<$ 8.0 deg., cloud W) are detected in dust, with a detection rate of 60\%, considerably higher than the lenticular detection rate of 41\% obtained on average for the Virgo cluster.
The evidence that dust may not be, at least in some cases, much affected by the diffuse hot gas might lead to a revision of our estimates about dust survival in passive ETG \citep{Cle10}.

Another mechanism which could produce the "incompatibility" between dust and atomic gas observed in Virgo ETG is the recombination of molecular hydrogen on the surface of cold dust grains \citep{Hol71}. If so, where dust is present, atomic hydrogen could recombine in molecules; the result is less atomic gas and more molecular gas in the dusty ETG. In fact in the only dust-detected ETG (VCC 1535), for which we have the mass of both the molecular hydrogen and the atomic hydrogen, their ratio is high ($M_{H_2}/M_{HI} \sim 30$), and there are several lower limits around 1 for this ratio in other ETG with dust. Nevertheless, dust favours recombination only locally, and this effect could not explain the lack of atomic hydrogen at great distances from the galaxy centre, since dust appears to be very concentrated in ETG. In the outer parts of galaxies the lack of hydrogen is probably caused by ram pressure stripping, which is quite effective on diffuse atomic gas, but not so effective on dust, particularly if it is concetrated and clumpy, owing to self shielding (see \citet{Cor10} for evidence that dust and gas can both be stripped in late-type Virgo galaxies). 

In summary, the presence in the Virgo cluster of several ETG with dust, but with only an upper limit to the \hi\ mass, can be explained by the different gas and dust survival times, by the stronger effects of ram pressure stripping on neutral atomic gas than on dust, and by the recombination of $H_2$ on grains. We are surprised, however, by the ETG for which \hi\ has been detected but for which there is only an upper limit to the dust mass (see Fig. 6). Among these, two are bright S0 galaxies with $M_B \simeq -19$: NGC 4262 and NGC 4270. The lack of dust in these objects could be explained by low metallicity, according to the relationship between the gas-to-dust mass ratio and the metallicity proposed by \citet{Dra07}. However, the metallicity of these two galaxies, although slightly below solar \citep{Kun10}, is not as low as the Draine relation would require to explain the observed upper limit on the dust-to-gas mass ratio. Both of these galaxies have been observed in CO by \citet{Com07}, but have not been detected.
We cannot exclude that some of these galaxies detected in \hi\ but not in dust might actually have dust emission whose peak does not coincide with the optical centre and, therefore, would be undetected by our procedure, which relies on the precise superposition of the optical and FIR peaks. This actually happens in the case of M87 (VCC881) and has been noticed, also thanks to the overlap of the dust emission with an H$\alpha$ filament. A similar procedure might also help to solve some of the other cases.

\section{Conclusions}

We have searched for dust in an optically-selected sample of 910 ETG in the Virgo cluster, 447 of which are complete to $m_{pg} \leq 18.0$, using the FIR images of HeViCS at 100, 160, 250, 350, and 500 $\mu$m, which cover a large fraction of the cluster.

From this study we obtain the following results:

1. We detect dust above the detection limit of 25.4 mJy at 250 $\mu$m in 46 ETG, 43 of which are in the optically complete part of the input sample. In addition, we detect dust at fainter levels in another 6 ETG. In all cases dust has been detected in more than one band.

2. We detect dust in the 4 ETG with synchrotron emission, including M 87. 

3. Considering only the complete detections (F$_{250}\geq 25.4$ mJy) out of the optically complete input sample, dust detection rates are 17.1\% for ellipticals, between 37.9\% and 41.4\% for lenticulars (S0 + S0a) and between 2.6\% and 3.2\% for dwarf ETG, depending on the effects of background galaxies. 

4. Dust appears to be much more concentrated than stars.

5. We estimate dust masses and temperatures with modified black-body fits. Dust masses range between $7 \times 10^4$ and $1.1 \times 10^7 M_{\odot}$, and temperatures between 14 and 31 K, with higher temperatures for more massive galaxies.

6. The dust mass does not correlate clearly with the stellar mass, and is often much greater than expected in a closed-box passively-evolving model, suggesting a possible external origin. 

7. In the Virgo cluster slowly rotating ETG appear more likely to contain dust than fast rotating ETG, contrary to what is observed for the molecular gas content for field galaxies, suggesting an environmental effect on the dust and molecular gas content in slow and fast rotators.

8. Comparing the dust results with those on \hi\ from ALFALFA, there are only 8 ETG detected both in dust and in \hi , while 39 have dust but no \hi , and 8 have \hi\ but no dust. Locations are also different, with the dusty ETG concentrated in the densest regions of the cluster, while the \hi\ rich ETG are at the periphery.

We conclude that, while general dust detection rates for ETG are smaller than for later galaxy types, the dust-to-stellar mass ratio for some dwarf ETG is comparable to that of the most dusty late-type galaxies. We suggest that at least in some cases the dust has an external origin in cluster ETG, and that the dust survives in ETG much longer than the neutral atomic gas.

\begin{acknowledgements}
We thank everyone involved with the Herschel Space Observatory and with SPIRE and PACS.
We thank Laura Ferrarese and Thorsten Lisker for useful information on the Virgo cluster galaxies and Bruce Draine for useful comments and advice.
S.diS.A., S.B., E.C., G.G., C.G., L.K.H., L.M., and C.P. are supported through the ASI-INAF agreements I/016/07/0 and I/009/10/0. C.P. is also supported by the PRIN-INAF 2009/11 grant. MG gratefully acknowledges support from the Science and Technology Foundation (FCT, Portugal) through the research grant PTDC/CTE-AST/111140/2009.
\end{acknowledgements}

\bibliography{Ref} 

\appendix 

\section{Revised coordinates for a subsample of the VCC}

\begin{figure}
\resizebox{\hsize}{!}{\includegraphics{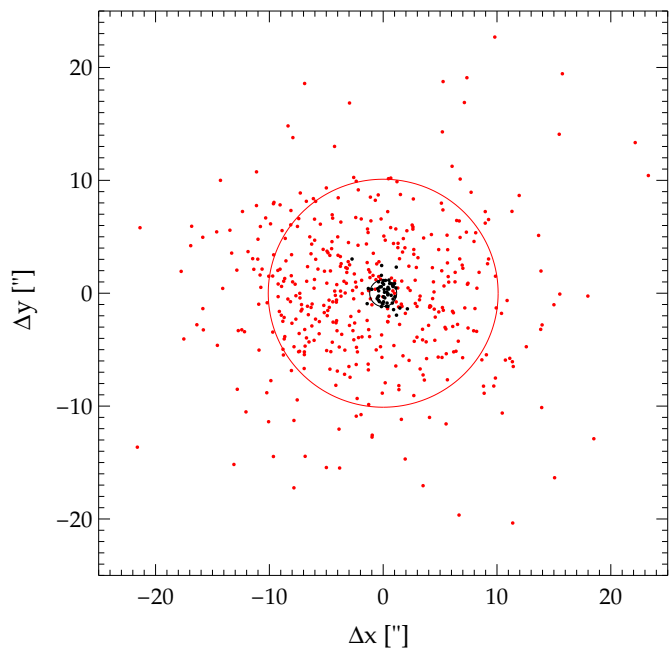}}
\caption{Offsets along R.A. and Dec. between the original coordinates and those of 
Table~\ref{vcc_rev} (red points). The red circle ($\Delta r = 10\arcsec$) contains 68\% 
of the sample considered here. Black dots and circle ($\Delta r = 1.2\arcsec$) show the offsets between the Lisker's coordinates (private communication) and our values.}
\label{fig_off}
\end{figure}

The coordinates for the VCC objects used in this work were originally extracted from NED.
For most objects more than two positions are available from the literature, and the 
coordinate accuracy quoted in NED is $\le 1\arcsec$.  However, for 476 VCC objects (both 
inside and outside the HeViCS fields), NED provides only the coordinates from the original work by \citet{Bin85} 
(as of September 2012); for these objects, NED quotes a positional accuracy of 25$\arcsec$ in both coordinates, 
which is inadequate for our work. Indeed, a visual check on SDSS7 images clearly showed that this 
inaccuracy of the position could result in an incorrect match with HeViCS sources.

Thus, we revised the coordinates by using SDSS7 $r$-band images, which are deeper and better suited to detect dwarf ETG. Thumbnail
images of $3\arcmin\times3\arcmin$ around the original coordinates were produced; the images
were smoothed, and contours overplotted as a guide to the eye; the object centre was then
flound manually by selecting a position on the image. For the fainter objects, we cross-checked
the identification of the source by using DSS-2 images and GOLDmine images, when available.
In the selection of the object which corresponds to the VCC galaxy, we were guided by the photographic magnitude $m_{pg}$ and the
dimensions indicated in the VCC catalogue.

We have been able to identify 454 objects, whose revised coordinates are given in Table~\ref{vcc_rev}.
For the remaining 22 VCC entries, instead, no unequivocal position could be found because there was
more than one suitable candidate object for a given position, or no identifiable object at all. These VCC galaxies are: 
VCC 203, 445*, 463*, 474*, 487*, 553*, 835, 852*, 910*, 913*, 927*, 969*, 987*,  
1116*, 1181, 1241*, 1260*, 1315, 1350*, 1571, 1835, 2079 (those with an asterisk are ETG in the HeViCS fields,
and are therefore part of the input sample for this paper). Unsurprisingly, most of them are dE with 
$m_{pg} > 19.0$. One notable exception is VCC 1571, whose position lies within 20$\arcsec$ of a relatively 
large ($\approx 30\arcsec \times 45\arcsec$) and bright ($m_{pg} =16.0$) dE,  VCC 1570, and is 
listed with very similar properties to those of its neighbour; this appears to be a duplicate 
entry in VCC. 

In Fig. A.1 we show the offsets along R.A. and Dec. between the original VCC positions and
the revised positions (red symbols). The red circle, which contains the offsets 
for 68\% of the objects, has a radius $\Delta r = 10\arcsec$. Assuming a Gaussian distribution of the offsets, 
this radius corresponds to the standard deviation. 
It corresponds to the coordinate accuracy quoted by \citet{Bin85}. While for most
of the objects $\Delta r \la 25\arcsec$ (the {\it NED accuracy}), for a few of the objects (20) the offset
is larger than that.
The largest offset is for VCC171, a relatively bright and large
Im galaxy ($m_{pg}=17.4$, size $\sim$ 35 x 20') which, thanks to the
illustrated atlas of \citet{San84} (where the galaxy
is called $8^\circ 7$), is identified with an object at $\Delta r = 3.25\arcsec$
from the original coordinates (beyond our original thumbnails).
Besides this galaxy, all other identified galaxies have $\Delta r \leq 67.5\arcsec$
(the value for VCC 709).
Offsets from
VCC coordinates, along with $m_{pg}$ and the GOLDMine morphological type, are given in Table~\ref{vcc_rev}.

Lisker et al. (2007) derived accurate coordinates for the Virgo cluster early-type dwarf galaxies with $m_{pg}\le 18$,
by minimising the object's asymmetry within a Petrosian aperture on SDSS5 images. For 53 of 
these objects, which are in common with those listed in Table~\ref{vcc_rev}, we show the offsets between the coordinates
obtained by Lisker (private communication) and those derived here (blue symbols) in Fig.~\ref{fig_off}. Despite
our {\em rough} manual technique, 68\% of our coordinates are within $1\arcsec$ (blue circle) of those
obtained by \citet{Lis07}, and all of them within $4\arcsec$. 

The offsets between the improved positions and the original positions by \citet{Bin85} are very similar
for our full set of 454 galaxies as for the 53 objects which we have in common with Lisker.
Thus, we believe that our position accuracy is within a few arcsec, also for the fainter galaxies 
($m_{pg} > 18$). We make this table available to the community as a valuable dataset, while waiting for 
the deeper imaging and more accurate positions which will be derived for
the Next Generation Virgo Cluster Survey \citep[NGVS,][]{Fer12}.

\section{The Virgo ETG detected in \hi\ with ALFALFA}

\begin{table*}
\caption{Virgo ETG detected in \hi\ in the $\alpha$.40 catalogue ($4^o<Dec<16^o$)}
\label{Table_B1}
\centering
\begin{tabular}{lllccrrcrcc}
\hline \hline
ID& Other & $m_{pg}$ & Type & Type & $cz_{opt.}$ & $cz_{HI}$ & D & $M_{HI}$ & $M_B$ & log($M_{HI}/L_B$) \\
~ & name &~ & GM\tablefootmark{1} & VCC/NED & $km/s$ & $km/s$ & Mpc & $10^7 M_\odot$ &~ & $M_\odot/L_\odot$ \\
\hline
CGCG69043 & NGC 4078 & 13.9 &  1 & S0?          & 2592 & 2572 & 17.0 &   4.8 & -17.37 & -1.46 \\
VCC 21   & IC 3025  & 14.75 & -3 & dS0(4)       &  506 &  485 & 17.0 &   5.9 & -16.49 & -1.02 \\
VCC 93   & IC 3052  & 16.3  & -1 & dE2          &  910 &  841 & 32.0 &  13.3 & -16.36 & -0.61 \\
VCC 94   & NGC 4191 & 13.57 &  2 & S0/a         &      & 2659 & 32.0 & 180.2 & -19.04 & -0.55 \\
VCC 180  &          & 15.3  &  1 & S0 pec       & 2232 & 2239 & 32.0 &   6.5 & -17.32 & -1.31 \\
VCC 190  &          & 18.0  & -1 & dE4          &      & 2352 & 32.0 &  13.8 & -14.63 &  0.09 \\
VCC 209  & IC 3096  & 15.15 & -3 & dS0?         & 1208 & 1263 & 17.0 &   3.9 & -16.14 & -1.06 \\
VCC 282  &          & 17.0  & -1 & dE5?         & 2014 & 1985 & 32.0 &  11.8 & -15.60 & -0.36 \\
VCC 304  &          & 16.3  & -1 & dE1 pec?     &  155 &  132 & 17.0 &   3.5 & -15.01 & -0.66 \\
VCC 355  & NGC 4262 & 12.41 &  1 & SB0          & 1359 & 1367 & 17.0 &  58.9 & -18.89 & -0.98 \\
VCC 375  & NGC 4270 & 13.11 &  1 & S0$_1$(6)    &      & 2377 & 32.0 &  49.9 & -19.50 & -1.30 \\
VCC 390  &          & 16.9  & -1 & dE3          & 2479 & 2474 & 32.0 &  21.2 & -15.71 & -0.15 \\
VCC 421  &          & 17.0  & -1 & dE2          &      & 2098 & 17.0 &   3.7 & -14.33 & -0.36 \\
VCC 710  &          & 14.9  & -3 & dS0:         & 1175 & 1182 & 17.0 &   7.8 & -16.41 & -0.86 \\
VCC 764  &          & 14.83 &  1 & S0$_2$(6)    & 2044 & 2020 & 17.0 &   6.8 & -16.33 & -0.89 \\
VCC 881  & NGC 4406 & 10.06 &  0 & S0$_1$(3)/E3 & -244 & -302 & 17.0 &   8.9 & -21.22 & -2.73 \\
VCC 956  &          & 18.75 & -1 & dE1,N:       &      & 2151 & 17.0 &  10.3 & -12.52 &  0.81 \\
VCC 1142 &          & 19.0  & -1 & dE           &      & 1306 & 23.0 &   9.2 & -12.89 &  0.62 \\
VCC 1391 &          & 18.5  & -1 & dE           &      & 2308 & 17.0 &   2.6 & -12.75 &  0.12 \\
VCC 1533 &          & 18.0  & -1 & dE2,N        &      &  648 & 17.0 &   2.9 & -13.25 & -0.02 \\
VCC 1535 & NGC 4526 & 10.61 &  1 & S0$_3$(6)    &  448 &  560 & 17.0 &   1.4 & -20.64 & -3.31 \\
VCC 1617 &          & 15.0  & -3 & d:S0(4) pec? &      & 1600 & 17.0 &   3.8 & -16.25 & -1.11 \\
VCC 1649 &          & 15.7  & -1 & dE3,N:       & 1038 &  972 & 17.0 &   1.4 & -15.54 & -1.28 \\
VCC 1993 &          & 15.3  &  0 & E0           &  875 &  925 & 17.0 &   5.0 & -15.96 & -0.88 \\
VCC 2062 &          & 19.0  & -1 & dE:          & 1146 & 1141 & 17.0 &  38.5 & -12.32 &  1.46 \\
CGCG100011 & NGC 4710 & 11.6  &  2 & SA(r)0+? sp  & 1129 & 1100 & 17.0 &   5.7 & -19.68 & -2.31 \\
\hline \hline
\end{tabular}
\tablefoot{GOLDMine type: -3=dS0 -2=dE$/$dS0 -1=dE(d:E) 0=E-E$/$S0 1=S0 2=S0a-S0/Sa}
\end{table*}

For a proper study of the cold ISM in ETG it is important to have information on the neutral atomic gas, in addition to the information on the dust discussed in this paper. The \hi\ survey ALFALFA \citep{Gio05,Gio07} has given us the important opportunity to survey the \hi\ content of Virgo galaxies in a complete and uniform way, as HeViCS is doing for the dust content. We published the first results on the \hi\ content of Virgo ETG from ALFALFA in a previous paper \citep[][hereafter dSA07]{diS07} in which we surveyed the 8-16 deg. declination strip, for which the \hi\ data were available at that time. Recently the ALFALFA survey has been extended to the 4-8 deg. declination strip \citep[][the so-called $\alpha$.40]{Hay11}, thereby covering the HeViCS area almost completely. Therefore we update here the list of \hi\ -detected ETG in the Virgo cluster for the 4-16 deg. declination strip (see Table~\ref{Table_B1}). The criteria of this \hi\ survey are the same as in dSA07, with the exception that we have now included the S0a-S0/SA galaxies (GOLDMine type = 2) among the surveyed ETG, in order to be consistent with what we have done for the HeViCS input ETG sample in the present paper. Actually, two such galaxies have been detected in \hi , one of which is in the 8-16 declination strip, already surveyed by dSA07. There is another change from dSA07: we are now using the distances as reported in GOLDMine, which are different for the various cluster components \citep{Gav99}. This has an obvious effect, even on distance dependent parameters, such as the luminosity and the derived \hi\ mass. The current $\alpha$.40 catalogue only contains sources detected with code 1 or 2 \citep[see][]{Hay11}. Therefore VCC 1964, which was detected with code 4 and was reported in dSA07, is not in the current list of \hi\ detected ETG.
The faint dE galaxy VCC 1202 deserves a special mention. We have remeasured its accurate position (see Appendix A), which is different from the position of the galaxy identified by \citet{Hay11} as the optical counterpart of the \hi\ source AGC 223724, although the latter is identified as VCC 1202 in their catalogue. The position of the optical counterpart of AGC 223724 given by \citet{Hay11} corresponds to a 18 mag galaxy to the south-west, which is indeed the most likely \hi\ counterpart. This galaxy is the faint blue galaxy No. 133 in \citet{Bor84}, but does not correspond to VCC 1202 of \citet{Bin85}, which is the much fainter and smaller galaxy for which we give an accurate position in Table~\ref{vcc_rev}. We have therefore excluded VCC 1202 from the list of Virgo ETG detected in \hi .

In summary, 26 Virgo ETG are detected in \hi , 22 of which are in the complete part of the optical sample ($m_{pg} \leq 18.0$), mostly based on the VCC . The sample of ETG examined in this update are all the galaxies present in GOLDMine, which have a GOLDMine type $-3 \leq T \leq 2$, which do not have a radial velocity $v_{hel} \geq 3000 km/s$, and which are in the 4-16 deg. declination strip. There are 1164 such galaxies, 575 of which are in the complete optical sample. In order to examine the \hi\ detection rate among Virgo ETG, we limit ourselves to this complete sample and we also take into account that ALFALFA has a much lower sensitivity in the region within one degree from M87, since this is a strong radio continuum source. Therefore, if we exclude the 59 ETG of the complete sample which are included in this M87 region, we have an \hi\ detection rate for Virgo ETG brighter than $m_{pg} = 18.0$ of 22 galaxies out of 516, i.e. 4.3\%. The fact that this rate is higher than the one obtained by dSA07 (2.3\%) is only marginally influenced by the fact that we have now included S0a-S0/SA galaxies, and is most likely thanks to the higher \hi\ detection rate in the outskirts of the Virgo cluster, as already remarked in dSA07, and has therefore been increased by the inclusion of the 4-8 deg. declination strip, which covers these outskirts. However the main conclusions of dSA07 and the conclusions drawn from the comparison with a sample of field ETG where the \hi\ detection rate is much higher \citep{Gro09}, are still valid.

\onecolumn

\onltab{
\begin{longtab}
\begin{longtable}{lrrrrrrrrr}
\caption{\label{vcc_rev} Revised coordinates}\\                   
\hline\hline       
VCC& \multicolumn{3}{c}{R.A. (J2000)}& \multicolumn{3}{c}{Dec. (J2000)}& offset & $m_{pg}$& type \\ 
& h & m & s&  d & $\arcmin$ & $\arcsec$ & \arcsec& &  \\ 
\hline     
\endfirsthead
\caption{continued.}\\
\hline\hline
VCC& \multicolumn{3}{c}{R.A. (J2000)}& \multicolumn{3}{c}{Dec. (J2000)}& offset & $m_{pg}$& type \\ 
& h & m & s&  d & $\arcmin$ & $\arcsec$ & \arcsec& &  \\ 
\hline
\endhead
\hline
\endfoot 
   2&12& 8&25.31&13&49&42& 1.3&18.5&-1\\
   8&12& 9&20.99&13&31&32& 4.2&19.0&-1\\
  11&12& 9&35.60& 6&44&35& 4.5&17.0&-1\\
  13&12& 9&45.84&13&33& 7&10.1&18.6&20\\
  23&12&10&24.94&13&21&58& 6.6&18.5&-1\\
  35&12&11&19.48&11&54&36& 6.3&19.0&20\\
  42&12&12& 5.99&14&57&11&10.0&19.5&-1\\
  44&12&12& 6.37& 9&53&47& 6.5&19.0&-1\\
  55&12&12&26.81&13&16&46& 3.5&18.7&-1\\
  63&12&12&41.39&10& 9&55& 5.8&19.3&-1\\
  69&12&12&55.57&10&33&35& 4.3&18.5&-1\\
  70&12&12&56.19&13& 4&11& 5.4&18.0&-1\\
  80&12&13&23.65& 7&45& 9& 8.3&20.0&-1\\
  85&12&13&35.84&13& 2& 4&10.2&17.5&20\\
  90&12&13&47.88&14&50&14&10.0&18.5&-1\\
  91&12&13&47.56& 6&21&46&11.4&18.0&-1\\
  98&12&13&54.16&13&52&20& 9.1&18.5&12\\
 100&12&14& 4.58&13&39& 8& 8.6&18.5&-1\\
 104&12&14& 8.87& 9&43&15& 6.2&18.0&19\\
 112&12&14&29.56&14& 4&36& 5.4&19.5&-1\\
 116&12&14&35.69& 7&15&13& 5.7&17.2&12\\
 129&12&15& 2.09&12&32&50& 8.7&14.6& 4\\
 133&12&15& 4.54&13& 6&29&17.4&19.5&-1\\
 136&12&15& 6.59& 4&47&45& 7.4&18.5&20\\
 139&12&15&12.32&11& 0& 1& 4.1&19.3&-1\\
 151&12&15&29.30&13&58& 6& 3.2&19.5&-1\\
 156&12&15&35.77&11&44&39& 2.2&19.5&-1\\
 164&12&15&53.02&12& 1&48& 6.4&20.0&-1\\
 171&12&16&12.04& 8&22&24& 195&17.4&12\\
 185&12&16&19.63&13& 8& 9& 8.3&19.3&-1\\
 188&12&16&22.52&14&15&24& 5.2&19.0&-1\\
 190&12&16&22.95& 7&47&55& 8.2&18.0&-1\\
 197&12&16&32.57&13& 9&50& 6.9&19.5&-1\\
 204&12&16&39.02&12&52&20& 4.1&18.5&20\\
 205&12&16&42.50& 6&41&21& 6.2&19.3&-1\\
 211&12&16&56.26& 4& 3&27& 8.2&17.5&-1\\
 214&12&16&57.69& 4&51&27& 5.0&18.2&-1\\
 219&12&17& 7.99& 7&21&12& 4.2&19.5&20\\
 229&12&17&18.56& 8&13& 7&11.0&20.0&-1\\
 232&12&17&24.68&13&30&39&25.0&19.0&-1\\
 239&12&17&30.69&10& 9&32& 8.9&20.0&-1\\
 240&12&17&31.26&14&21&21& 1.4&19.3&-1\\
 242&12&17&36.29& 5&25&29& 8.9&18.5&-1\\
 244&12&17&39.73& 5&20&29& 3.6&18.0&-1\\
 247&12&17&40.40& 8&23&12& 7.6&18.0&20\\
 250&12&17&45.55& 6&25&51& 5.5&19.5&-1\\
 252&12&17&46.00& 5&26&53& 3.1&17.3&-1\\
 253&12&17&46.30& 3&51&47& 6.0&19.0&-1\\
 254&12&17&46.05& 6&25&24& 5.5&20.0&-1\\
 255&12&17&46.60& 7&15&44& 8.0&18.5&12\\
 274&12&18& 7.53& 5&55&49&52.3&17.5&17\\
 276&12&18&12.37& 5&39& 8& 6.5&18.4&20\\
 285&12&18&26.99&10&39& 2& 3.0&19.5&-1\\
 290&12&18&30.58&12&23&57& 7.7&20.0&-1\\
 291&12&18&31.38& 4&58&44& 1.8&18.8&-1\\
 294&12&18&35.26& 6&29&39& 6.9&18.0&-1\\
 296&12&18&37.31& 6&18&47& 2.4&19.3&-1\\
 306&12&18&48.35& 8&59&17& 9.6&19.3&-1\\
 310&12&18&53.82&12&11&38& 3.3&19.5&-1\\
 316&12&19& 1.19& 5&46&23&63.2&19.3&20\\
 317&12&19& 2.70& 5& 5&51& 8.9&18.0&-1\\
 320&12&19& 5.13& 4&39&35& 9.0&16.5&18\\
 326&12&19&11.52& 6&29&34& 1.8&20.0&-1\\
 360&12&19&36.86&15&27&17&15.2&18.5&-1\\
 365&12&19&44.57& 3&50&40&14.3&18.7&-1\\
 367&12&19&45.30& 5&27&21& 9.1&17.2&12\\
 368&12&19&44.86& 7&37&49&12.0&19.0&-1\\
 372&12&19&47.65&14&42&22& 9.5&18.0&-1\\
 379&12&19&51.64& 5&59&46&11.1&17.0&20\\
 381&12&19&53.59& 6&39&56& 7.7&16.5&12\\
 383&12&19&57.24& 9&33&32& 6.7&20.0&-1\\
 384&12&19&58.60& 6&20& 3& 8.3&20.0&-1\\
 387&12&20& 3.57& 5&43&13& 7.9&18.5&-1\\
 399&12&20&14.20& 5&54&56& 0.3&19.0&-1\\
 403&12&20&17.55&10&19&15& 9.3&18.0&-1\\
 405&12&20&17.87& 6& 0& 6&10.6&20.0&-1\\
 412&12&20&24.51&10&40&53&14.8&20.0&-1\\
 414&12&20&24.37&14&41&28&10.9&17.9&19\\
 416&12&20&26.83&12&47&50&14.6&20.0&19\\
 418&12&20&26.86&14&47& 7& 8.2&17.9&-1\\
 425&12&20&35.73& 8&12& 4& 4.9&17.3&12\\
 427&12&20&41.20& 4&47&49&10.5&18.5&-1\\
 431&12&20&46.37&12&45&27& 7.8&18.5&-1\\
 440&12&20&52.21& 5&13&55& 6.3&17.2&-1\\
 441&12&20&52.85& 6&19&28&12.5&18.5&20\\
 444&12&20&54.68&14&59&22&11.7&17.2&-1\\
 456&12&21& 9.06&12&17&58& 8.2&19.5&-1\\
 457&12&21& 9.88& 6&22& 8& 7.6&19.0&-1\\
 467&12&21&19.27& 3&47&16& 5.0&17.7&12\\
 476&12&21&29.11&10&29&10&10.9&17.9&12\\
 477&12&21&27.10&15& 1&12&20.1&17.0&12\\
 480&12&21&29.24&12&47&52& 8.2&19.5&-1\\
 481&12&21&30.81&15&30& 3& 9.1&18.3&-1\\
 484&12&21&33.94& 6&32&49&17.4&19.2&-1\\
 496&12&21&42.73& 9&21&19& 5.2&18.7&-1\\
 499&12&21&44.56& 9&14&12&12.8&17.6&-1\\
 502&12&21&48.83&11&51& 6& 7.6&19.5&-1\\
 506&12&21&54.53& 4&45&12&14.2&18.5&-1\\
 507&12&21&52.30&15&22&56& 8.0&19.0&-1\\
 511&12&21&55.83& 8&20&45& 9.6&18.2&-1\\
 518&12&22& 0.21&11&38& 8& 7.2&20.0&20\\
 532&12&22&10.46&11&38&29& 7.8&18.7&-1\\
 537&12&22&12.78&12&41&48&10.4&18.5&-1\\
 540&12&22&16.90&10&57&12& 3.1&19.4&-1\\
 548&12&22&22.90& 3&44&39& 7.2&18.3&12\\
 555&12&22&27.55& 7& 5&52& 0.8&19.2&-1\\
 563&12&22&35.27&16&20&52& 6.1&16.3&-1\\
 564&12&22&36.50&11&17&50&17.1&20.0&-1\\
 569&12&22&39.05&10&37&43& 9.4&19.5&-1\\
 577&12&22&44.62& 5&24&55& 9.0&18.3&-1\\
 579&12&22&45.33& 7&15&51& 2.6&20.0&-1\\
 581&12&22&46.55& 7& 6& 4& 3.8&20.0&20\\
 582&12&22&47.24& 8&26&14&10.2&20.0&-1\\
 585&12&22&46.97&11&20&40&18.9&17.0&12\\
 589&12&22&51.18& 7& 4&13& 4.7&19.8&-1\\
 590&12&22&51.04& 8&54& 8& 6.0&19.0&-1\\
 591&12&22&51.85& 7& 7&59& 5.7&20.0&-1\\
 595&12&22&53.72&12&23&19& 9.2&19.5&-1\\
 609&12&23& 3.46& 8&32& 7& 8.7&19.2&20\\
 610&12&23& 3.17&16&19&41& 4.7&20.0&-1\\
 612&12&23& 5.32& 4& 7&48& 6.7&17.3&20\\
 614&12&23& 4.17&14&48&21&14.6&18.6&-1\\
 628&12&23&16.15& 7&41&13&12.7&18.2&12\\
 629&12&23&17.67& 6&54&43& 6.0&18.2&12\\
 631&12&23&18.46& 7&39&46& 5.1&18.5&-1\\
 633&12&23&21.43& 7&35&30& 8.6&18.0&-1\\
 639&12&23&25.19& 6&59&54& 3.7&19.0&-1\\
 642&12&23&30.33& 5&29&41& 5.4&18.3&-1\\
 645&12&23&31.88&11&16&14& 8.0&18.2&-1\\
 653&12&23&36.91& 7&35&44& 6.8&17.6&-1\\
 658&12&23&39.33&10& 1&51& 2.6&18.5&-1\\
 659&12&23&38.54&12&37&39& 5.3&18.3&-1\\
 661&12&23&41.31& 7&17&27& 7.9&18.5&-1\\
 666&12&23&46.24&16&47&26& 2.1&16.8&12\\
 671&12&23&52.81& 5&45&19& 3.5&18.0&-1\\
 676&12&23&54.20& 6&54& 2& 5.1&19.5&-1\\
 678&12&23&52.94&12&46&22&22.1&18.3&-1\\
 686&12&23&59.83& 9&29&37& 4.7&18.5&-1\\
 689&12&23&59.92&17&39& 5& 5.8&19.8&20\\
 691&12&24& 0.63&11&51&34& 8.4&19.5&-1\\
 694&12&24& 5.80& 7&31&53& 7.4&18.8&-1\\
 701&12&24& 8.25&11& 8&45& 4.8&19.2&-1\\
 702&12&24& 9.05& 8&30&49& 4.5&19.5&-1\\
 705&12&24&10.94&11&56&47&14.1&17.2&-1\\
 707&12&24&12.92&11&45&41& 2.7&19.0&-1\\
 709&12&24&12.73&14&29&36&67.5&19.0&-1\\
 716&12&24&13.68&14&55&44&13.8&19.2&-1\\
 724&12&24&25.63& 7& 7&51& 1.4&18.3&-1\\
 730&12&24&27.99& 6&40&36&16.0&20.0&-1\\
 732&12&24&30.96&11&48&41&12.0&18.5&-1\\
 743&12&24&42.50&11&28&56&13.5&20.0&-1\\
 744&12&24&46.84& 7&55& 3& 8.0&19.0&-1\\
 746&12&24&47.67& 8&26&19& 2.8&17.7&-1\\
 752&12&24&47.90&11&49& 6&31.3&19.0&-1\\
 769&12&25& 4.23&15&42&40&17.8&17.2&-1\\
 774&12&25&10.05&10&27&23& 7.3&18.5&-1\\
 776&12&25&11.15& 6&18&51&25.5&18.2&-1\\
 783&12&25&15.57& 7&14&18& 8.0&19.0&-1\\
 791&12&25&22.21& 6&42&40& 7.3&16.4&-1\\
 803&12&25&28.97&12&29&37&12.7&18.0&-1\\
 807&12&25&33.18& 7&48&28& 7.6&19.0&-1\\
 811&12&25&38.23&10&15& 0&31.7&16.5&19\\
 814&12&25&36.67&12&50&59& 3.3&19.0&20\\
 824&12&25&38.70&14& 9& 2&11.5&18.0&-1\\
 842&12&25&48.55&12&13&32& 7.0&20.0&-1\\
 853&12&25&55.75&11&48& 4&28.1&20.0&-1\\
 875&12&26& 9.87& 7&18& 8& 7.4&19.8&20\\
 879&12&26&12.30& 6& 5&10& 6.4&19.0&-1\\
 880&12&26&12.13&12& 5&10& 6.4&19.6&-1\\
 883&12&26&15.47& 7&45&21& 7.2&19.0&-1\\
 884&12&26&15.34&13& 8&35& 5.9&18.5&-1\\
 886&12&26&15.29&13&20&26& 8.2&20.0&-1\\
 895&12&26&24.92&10&34&53&33.5&18.3&-1\\
 896&12&26&22.58&12&47& 9&10.0&18.0&-1\\
 900&12&26&25.99&13&44&18& 9.9&18.5&-1\\
 901&12&26&26.12&16&31&23& 1.2&18.0&-1\\
 902&12&26&28.65& 8&47&43&13.4&19.1&-1\\
 903&12&26&28.07&12&55&13&13.4&18.9&-1\\
 906&12&26&29.85& 9&58&44&20.9&19.0&-1\\
 914&12&26&34.28& 8&59&34& 7.3&19.0&-1\\
 922&12&26&36.98&10&12&28&25.5&18.2&-1\\
 923&12&26&36.36&12&48&10& 4.5&19.5&-1\\
 924&12&26&34.90&13&52&13&18.0&19.3&-1\\
 925&12&26&38.06&15& 5& 7&17.2&19.5&-1\\
 930&12&26&41.11&12&50&43&11.5&18.0&-1\\
 937&12&26&46.58&13&15&59&13.1&19.0&-1\\
 943&12&26&47.58&13&40&47& 9.6&18.6&-1\\
 948&12&26&53.28& 7&45& 9& 6.9&20.0&-1\\
 954&12&26&56.31& 5&58&18&12.3&17.2&-1\\
 960&12&26&58.80& 6&48& 3& 6.6&20.0&-1\\
 964&12&27& 1.91&14& 6&31& 8.7&18.4&-1\\
 967&12&27& 3.81&12&51&58& 4.7&18.7&-1\\
 968&12&27& 6.04&13&19&24& 5.7&19.2&-1\\
 970&12&27& 7.97& 7&48&59&12.3&19.5&-1\\
 982&12&27&12.73&17&49&15& 2.5&17.3&-1\\
 983&12&27&15.16& 9&37&34&24.7&18.0&-1\\
 988&12&27&17.67& 6&46& 2& 7.3&20.0&-1\\
 993&12&27&20.55& 9&35&29&22.6&18.5&-1\\
 999&12&27&24.11&12& 7&59&12.8&19.7&-1\\
1000&12&27&24.73&11&14&17& 8.0&18.2&-1\\
1004&12&27&24.90&13&24&23&15.6&19.0&-1\\
1006&12&27&25.24&14&26& 1&10.9&18.2&-1\\
1009&12&27&27.03& 8&21& 8&11.6&19.5&-1\\
1012&12&27&28.46& 8&23&28&17.4&20.0&-1\\
1014&12&27&29.77&12&15& 5& 1.0&18.3&-1\\
1023&12&27&34.38&12&48&13&12.0&20.0&-1\\
1027&12&27&39.23&12&52&47&18.0&18.1&-1\\
1029&12&27&39.02&14&32&31& 8.7&19.0&-1\\
1032&12&27&42.65& 6&22& 6& 2.2&19.5&12\\
1038&12&27&41.82&14&35&55&11.7&18.6&20\\
1040&12&27&44.53&12&59& 1& 6.5&17.5&-1\\
1041&12&27&46.50&11&44&28&12.1&19.5&20\\
1045&12&27&49.64& 7& 1&33&17.0&18.7&12\\
1046&12&27&49.43&12&29&59&10.2&20.0&-1\\
1050&12&27&55.16& 8&50&53& 2.7&18.5&-1\\
1051&12&27&54.55&12&36&15& 4.5&19.8&-1\\
1053&12&27&54.79&13&49&15&10.1&19.5&-1\\
1054&12&27&56.02& 8& 5&15& 9.5&18.5&-1\\
1056&12&27&57.33&14&28&13& 2.6&18.7&-1\\
1067&12&28& 6.31& 8& 3&42& 7.3&20.0&20\\
1070&12&28& 6.73&12&58&43&11.3&19.6&-1\\
1071&12&28& 7.21& 8&48&54& 5.9&18.4&-1\\
1077&12&28&10.24&12&48&31& 7.1&19.2&-1\\
1081&12&28&12.85&13& 0&54&10.9&18.8&-1\\
1082&12&28&14.32& 6&18&40& 9.7&20.0&20\\
1089&12&28&17.65&10&52& 6&15.5&17.9&-1\\
1090&12&28&17.84& 9&43&40& 7.1&19.0&-1\\
1092&12&28&20.08& 9&10&12& 6.2&17.0&-1\\
1094&12&28&20.68& 6&11&43& 3.4&17.3&20\\
1097&12&28&20.25&15&41&59& 8.2&19.0&-1\\
1098&12&28&22.39& 8&43&41& 4.6&18.5&12\\
1106&12&28&31.47&10&31& 7&32.5&17.5&12\\
1108&12&28&29.78& 8&32&16& 5.8&19.5&-1\\
1109&12&28&31.09& 6&54&16& 8.3&18.5&-1\\
1112&12&28&31.46&16& 3&51& 2.7&18.8&-1\\
1113&12&28&34.03& 7&34&17& 4.5&19.5&-1\\
1115&12&28&32.42&11&44&39&13.0&17.7&-1\\
1117&12&28&40.00& 8&49&37& 4.5&19.0&-1\\
1121&12&28&41.08&11& 7&57&10.9&16.5&12\\
1124&12&28&44.10&10&51&56&20.5&16.3&-1\\
1128&12&28&44.95& 9& 3&15& 4.5&17.3&12\\
1133&12&28&48.47& 8&13&53& 5.3&19.3&-1\\
1137&12&28&49.68&14& 9&26&17.9&17.2&-1\\
1140&12&28&52.63&14&23&38& 9.8&18.5&-1\\
1151&12&28&59.22& 7&31&35& 7.3&16.7&-1\\
1159&12&29& 4.28& 4&50&55& 9.3&19.0&-1\\
1160&12&29& 4.25& 8&27&21&10.3&18.3&12\\
1165&12&29&10.28& 9&16& 0& 1.4&17.9&19\\
1167&12&29&14.69& 7&52&39&37.2&15.9&-1\\
1170&12&29&13.03&10&59&26& 7.2&19.0&-1\\
1171&12&29&14.60& 8&14&29& 9.2&19.2&-1\\
1176&12&29&18.96& 8& 1&11& 6.8&19.0&20\\
1180&12&29&21.20&16&48&22& 8.7&16.5&-1\\
1194&12&29&29.29&14&10&10& 9.2&18.3&-1\\
1195&12&29&31.91& 7&49&18& 6.4&19.0&19\\
1201&12&29&34.51&13&19&56&13.7&18.8&-1\\
1202&12&29&35.58&13&12&39& 8.8&20.0&-1\\
1207&12&29&37.95& 9&31&15& 3.3&17.5&-1\\
1211&12&29&39.64& 9&27&48& 5.9&18.2&-1\\
1212&12&29&39.02&11&38& 0&10.6&16.9&-1\\
1214&12&29&38.41&14& 3&30&16.6&20.0&-1\\
1215&12&29&40.45&16&57&42& 8.5&19.4&-1\\
1220&12&29&42.96&14&22& 4& 9.5&19.0&20\\
1221&12&29&43.41&17&30&55& 4.4&18.2&-1\\
1227&12&29&46.98&11&10&12&18.3&17.9&19\\
1228&12&29&45.12&14&37&52&13.2&16.2&-1\\
1229&12&29&47.15&13& 4&34& 5.2&19.4&20\\
1230&12&29&49.50& 7&17& 1& 5.1&19.5&12\\
1232&12&29&49.30&11&29&36& 4.6&19.0&20\\
1236&12&29&51.98& 6&11&39& 4.7&18.4&-1\\
1244&12&29&56.35&13&13&12& 9.2&18.9&-1\\
1246&12&29&59.40&10&51&21&25.9&18.2&-1\\
1248&12&29&57.64&14&28&21& 5.5&18.2&-1\\
1251&12&30& 1.13&13& 7& 4&11.8&19.8&-1\\
1255&12&30& 6.85& 6&20&36&12.2&19.0&12\\
1265&12&30&11.56&13&41&28&16.2&19.0&-1\\
1267&12&30&13.34& 7& 6& 9& 5.3&19.3&-1\\
1275&12&30&17.73& 7&53& 2& 8.8&20.0&-1\\
1276&12&30&18.14&14&40&51&22.2&20.0&-1\\
1280&12&30&17.20&14& 7&45& 9.6&19.5&20\\
1281&12&30&18.78& 7&54&19& 9.2&19.5&-1\\
1282&12&30&18.21&12&34&17&58.1&19.7&-1\\
1285&12&30&20.31&14& 8&24&15.9&18.5&-1\\
1296&12&30&32.71& 6&31&56& 1.3&17.2&-1\\
1301&12&30&38.22&13&37&10& 8.6&19.0&-1\\
1306&12&30&45.88& 9& 0&42& 4.3&19.5&-1\\
1310&12&30&46.88&13&12&49& 6.4&19.3&-1\\
1312&12&30&47.27&11&32&15&29.0&18.7&-1\\
1319&12&30&52.31&13&51&30& 4.7&19.5&-1\\
1325&12&30&56.01&13&26&55& 4.2&20.0&-1\\
1328&12&30&57.37&13&37&12& 9.3&19.3&-1\\
1329&12&31& 0.56& 5&33&18&12.4&18.1&-1\\
1332&12&31& 2.32& 6& 7&20&12.3&18.3&19\\
1336&12&31& 3.96&11&50&11& 6.2&17.0&19\\
1337&12&31& 4.84&15& 4&12&15.0&18.0&-1\\
1338&12&31& 4.64&17&23&16& 9.5&18.7&-1\\
1342&12&31&13.79& 6&40&16& 4.9&19.0&-1\\
1344&12&31&14.22&16&57& 4& 8.1&20.0&-1\\
1345&12&31&15.83& 9&21&32& 7.4&19.0&-1\\
1349&12&31&17.32& 7&51&40& 4.3&18.5&-1\\
1357&12&31&24.42& 9&28&28& 4.0&18.9&12\\
1361&12&31&27.13& 9&43&59& 6.2&17.2&-1\\
1364&12&31&28.86& 3&58&28&12.5&19.5&-1\\
1365&12&31&30.73&10& 0&19& 5.2&19.3&-1\\
1367&12&31&33.10& 8&58& 4&28.4&19.5&12\\
1370&12&31&36.88&11& 0&26&31.6&17.4&-1\\
1371&12&31&35.79&13&49&27& 6.6&18.2&-1\\
1372&12&31&35.87&16&43&29& 8.9&19.5&-1\\
1378&12&31&41.67& 5&46&27& 9.0&18.2&-1\\
1382&12&31&44.58&10& 0&46& 3.7&19.5&20\\
1383&12&31&46.63& 9& 5& 1&12.5&18.5&20\\
1384&12&31&46.63&10&40&27&19.8&17.1&-1\\
1387&12&31&50.53&13&40&50&12.7&19.2&-1\\
1388&12&31&49.91&16&58&18& 4.9&19.0&12\\
1390&12&31&51.83&14&22&20&12.6&20.0&-1\\
1391&12&31&53.82& 5&10&22& 2.2&18.5&-1\\
1397&12&31&59.98& 3&32&24& 6.9&18.0&12\\
1403&12&32& 0.25&13& 4&52& 7.9&17.1&12\\
1404&12&32& 1.51& 8&40&10& 6.2&18.1&-1\\
1406&12&32& 2.77& 8& 4&30& 4.9&19.6&-1\\
1408&12&32& 4.41& 7&56&37& 2.0&18.0&19\\
1415&12&32& 8.81& 8& 2&21& 4.3&19.5&-1\\
1421&12&32&13.00& 9&18&54& 2.8&20.0&-1\\
1424&12&32&20.56&10&18&37& 8.4&19.5&20\\
1425&12&32&20.78&10& 3&26& 5.8&18.9&-1\\
1433&12&32&28.46&14&21&45&13.7&19.0&-1\\
1434&12&32&29.82&14&39&44&36.9&20.0&-1\\
1439&12&32&35.23& 8&38&10& 6.6&20.0&-1\\
1447&12&32&38.66&10&45&34&15.9&19.5&-1\\
1467&12&32&55.00&13&53&19& 7.4&18.8&-1\\
1470&12&33& 0.40&10& 7&42& 3.4&20.0&-1\\
1477&12&33& 6.41&13&18&10&11.3&19.0&-1\\
1478&12&33& 6.22&15&30&27& 7.5&18.3&-1\\
1487&12&33&12.28& 8&22&39& 8.2&19.5&-1\\
1505&12&33&24.64&15&24&27&10.6&18.0&-1\\
1506&12&33&27.96& 7&51&16& 5.4&18.3&-1\\
1510&12&33&30.14&16&17&43& 5.0&18.3&12\\
1511&12&33&33.21&13&28&36& 9.9&18.2&-1\\
1515&12&33&41.11& 3&40&33& 3.2&17.0&20\\
1520&12&33&42.25&10&47&14& 6.8&20.0&-1\\
1527&12&33&50.52&10&59&14& 9.9&19.8&-1\\
1533&12&34& 1.47& 5&57&11& 1.5&18.0&-1\\
1534&12&34& 1.60& 6&38&42& 4.9&18.2&-1\\
1543&12&34& 9.67&16&42&42& 4.0&18.0&-1\\
1546&12&34&11.28&13&34&22& 6.9&19.5&-1\\
1547&12&34&14.87& 2&17&47& 1.5&18.2&-1\\
1553&12&34&16.13&16& 3&31&14.8&16.7&-1\\
1556&12&34&18.61&14&28&23&10.2&19.3&-1\\
1560&12&34&23.04&10&57&48&10.6&20.0&20\\
1580&12&34&43.47& 8& 5&26& 5.1&18.8&-1\\
1586&12&34&46.51&14&17&32& 6.1&19.3&-1\\
1589&12&34&53.49& 8& 9&36& 6.2&19.0&-1\\
1590&12&34&54.96&14&39& 0& 3.6&20.0&-1\\
1591&12&34&57.15&10& 3&46& 9.7&18.5&-1\\
1600&12&35& 6.96&10&30&35& 7.1&19.0&-1\\
1601&12&35& 8.11&12&57&56&12.5&18.0&-1\\
1612&12&35&24.41&13&45&56& 5.0&18.5&-1\\
1618&12&35&31.68& 7& 1&33& 9.5&18.3&-1\\
1622&12&35&31.40&16& 2&53&15.2&17.9&-1\\
1640&12&35&49.68& 9&20&31& 9.5&19.0&-1\\
1645&12&35&55.23& 2&29&10& 4.2&18.0&-1\\
1646&12&35&57.41& 5&47&23& 9.0&19.5&-1\\
1648&12&36& 2.51& 6&24&28& 6.9&20.0&-1\\
1650&12&36& 2.96&13&29& 8& 7.4&17.4&-1\\
1651&12&36& 7.19& 6& 3&15& 4.9&17.0&-1\\
1652&12&36& 8.56& 7&35&46&12.9&17.3&-1\\
1655&12&36&14.18& 9&28&18& 5.1&18.6&-1\\
1666&12&36&29.40& 8&30&44& 3.7&19.5&-1\\
1693&12&36&52.54&13&17& 0&17.4&19.2&-1\\
1694&12&36&52.60&13&46& 3& 4.1&19.3&-1\\
1702&12&37& 6.96&13&58&54&10.8&17.7&-1\\
1703&12&37& 9.23& 8&23&31& 3.1&19.0&20\\
1705&12&37&10.84& 9&29&24& 9.1&17.8&-1\\
1707&12&37&15.11&10&43&13& 2.0&19.0&-1\\
1714&12&37&25.24&14&18&44& 7.8&18.5&-1\\
1716&12&37&27.09&15& 8&56& 6.3&19.0&-1\\
1719&12&37&31.17& 9&30&20& 6.4&18.5&-1\\
1723&12&37&36.29& 9&12&24& 1.4&20.0&-1\\
1724&12&37&41.08& 7&40&37& 5.9&18.5&-1\\
1731&12&37&49.11&13&26&24& 5.5&18.5&-1\\
1732&12&37&50.26&15&43&38&14.0&17.8&-1\\
1733&12&37&52.17&14&28& 1&12.5&18.0&-1\\
1734&12&37&54.04& 2& 8&56& 6.1&20.0&-1\\
1735&12&37&52.62& 2&10&50&25.6&19.5&-1\\
1738&12&37&54.74&15& 9&32& 1.8&18.3&-1\\
1739&12&37&57.71& 6&32&31&10.6&18.7&-1\\
1741&12&38& 2.77& 2&22&10& 8.4&18.4&-1\\
1742&12&38& 4.00& 9&31&26& 4.5&19.0&-1\\
1749&12&38&12.13&10&42& 7& 8.2&19.5&-1\\
1751&12&38&13.56&15&22&25& 2.4&18.6&-1\\
1753&12&38&16.01&14&52&12& 1.9&16.8&12\\
1766&12&38&37.46&13&30&28&14.2&18.5&-1\\
1770&12&38&42.68& 6&40& 7&10.5&18.4&20\\
1771&12&38&43.88& 9&31&56& 1.3&18.0&12\\
1775&12&38&53.09&12&59& 0& 6.7&20.0&-1\\
1776&12&38&54.09&10&14&34&16.6&17.1&19\\
1784&12&39&14.18&15&37&38&12.8&15.8&12\\
1786&12&39&16.31&12&58&15&14.9&18.1&-1\\
1788&12&39&18.54&10&49& 9& 6.8&19.7&-1\\
1792&12&39&25.98&12&55&41& 4.4&18.3&-1\\
1797&12&39&33.67& 2&42&40& 4.3&18.3&-1\\
1801&12&39&39.63& 2&13&58& 5.2&19.5&-1\\
1805&12&39&42.69& 9&18&34& 9.3&19.5&-1\\
1819&12&40& 5.17& 5&45&32& 6.6&18.8&20\\
1820&12&40& 4.17&13& 2& 3& 9.2&19.0&12\\
1840&12&40&26.93&10& 7&41& 2.6&18.5&-1\\
1844&12&40&29.20& 9&43&22& 3.9&19.5&20\\
1850&12&40&34.13&15&22& 2&12.8&19.5&20\\
1851&12&40&43.60&10&24&28&10.1&18.8&-1\\
1852&12&40&43.69&13&47&22& 3.0&19.5&-1\\
1856&12&40&50.98& 7&55&49& 4.3&18.8&-1\\
1864&12&41& 6.88& 3&36&38& 4.8&16.8&20\\
1867&12&41&10.66&15& 2&55&20.8&17.7&-1\\
1872&12&41&19.04& 2& 6& 0&16.0&18.5&-1\\
1874&12&41&18.12&13& 9&54&15.4&17.7&-1\\
1877&12&41&23.95& 8&21&56& 3.7&18.6&-1\\
1884&12&41&39.36& 9&12&33&19.5&16.7&19\\
1885&12&41&38.39&15&49&12&23.3&16.4&12\\
1899&12&41&57.45& 8&49&52& 7.7&20.0&-1\\
1900&12&41&57.24&13& 4& 5&13.6&16.6&19\\
1907&12&42& 5.99& 4&12&18&12.1&19.0&20\\
1911&12&42& 9.74& 4&41&30&16.0&19.5&-1\\
1914&12&42&11.36&14&49& 2&27.5&19.5&20\\
1919&12&42&18.21&10&34& 3&10.2&17.0&-1\\
1926&12&42&34.03&10&20&36&16.1&19.3&-1\\
1928&12&42&37.83&13&36&23& 6.4&17.5&-1\\
1930&12&42&41.50& 7&50&37&12.4&19.5&-1\\
1970&12&43&29.04&10& 5&35& 2.4&15.8&12\\
1974&12&43&37.05& 3&21&45&15.5&19.0&-1\\
1975&12&43&35.42&10&45&20&11.9&18.8&-1\\
1976&12&43&35.81&13&15& 2& 4.4&18.5&-1\\
1994&12&44&13.19& 9&43&23& 6.1&16.8&19\\
1996&12&44&20.89& 8&24&13& 3.8&18.5&-1\\
2013&12&45& 8.49& 8&35&22& 1.8&18.3&-1\\
2014&12&45& 9.71&10&37&49& 8.8&17.5&-1\\
2015&12&45&11.96&10&19&28& 5.5&16.2&17\\
2021&12&45&27.71&13& 1&36& 7.5&19.0&-1\\
2022&12&45&33.11& 8&43&45& 4.0&20.0&-1\\
2026&12&45&34.66&14&23&13& 0.9&20.0&-1\\
2027&12&45&36.91&10&35&56&15.3&18.5&12\\
2030&12&45&50.72& 8&18&22& 4.8&18.8&-1\\
2031&12&45&49.63&14&31&17& 1.4&19.5&-1\\
2035&12&46& 9.44& 8&42&28& 3.6&20.0&-1\\
2052&12&47&29.46&13&38&56& 6.4&20.0&-1\\
2053&12&47&31.35& 7&56& 3& 1.5&19.2&-1\\
2059&12&47&48.43& 7&42&22& 9.3&20.0&-1\\
2060&12&47&55.50& 9& 9&32&10.5&18.6&-1\\
2063&12&48& 3.23&13&43&27& 8.1&17.0&-1\\
2067&12&48&20.87& 8& 2&28&10.9&18.2&-1\\
2068&12&48&20.51&10& 1&39&10.3&18.8&-1\\
2075&12&48&29.39&13&48&18& 6.4&19.3&-1\\
2084&12&50&24.00&10&29&25& 8.9&19.5&-1\\
2091&12&52& 5.91&11& 1&56& 6.1&18.5&-1\\
2094&12&52&35.64&10&26&58&10.2&17.8&12\\

\hline \hline
\end{longtable}
\end{longtab}
}



\end{document}